\documentclass{article}%
\usepackage{amssymb}
\usepackage{amsmath}
\usepackage{graphicx}
\usepackage{amsfonts}%
\providecommand{\U}[1]{\protect\rule{.1in}{.1in}}

\newtheorem{theorem}{Theorem}

\newtheorem{example}[theorem]{Example}

\begin{document}

\title{Partitions and Objective Indefiniteness\\in Quantum Mechanics}
\author{David Ellerman\\Department of Philosophy \\U. of California/Riverside}
\maketitle

\begin{abstract}
Classical physics and quantum physics suggest two meta-physical types of
reality: the classical notion of a objectively definite reality with
properties "all the way down," and the quantum notion of an objectively
indefinite type of reality. The problem of interpreting quantum mechanics (QM)
is essentially the problem of making sense out of an objectively indefinite
reality. These two types of reality can be respectively associated with the
two mathematical concepts of subsets and quotient sets (or partitions) which
are category-theoretically dual to one another and which are developed in two
mathematical logics, the usual Boolean logic of subsets and the more recent
logic of partitions. Our sense-making strategy is "follow the math" by showing
how the logic and mathematics of set partitions can be transported in a
natural way to Hilbert spaces where it yields the mathematical machinery of
QM--which shows that the mathematical framework of QM is a type of logical
system over $%
\mathbb{C}
$. And then we show how the machinery of QM can be transported the other way
down to the set-like vector spaces over $%
\mathbb{Z}
_{2}$ showing how the classical logical finite probability calculus (in a
"non-commutative" version) is a type of "quantum mechanics" over $%
\mathbb{Z}
_{2}$, i.e., over sets. In this way, we try to make sense out of objective
indefiniteness and thus to interpret quantum mechanics.

\end{abstract}
\tableofcontents

\section{Two types of reality}

\subsection{Objective definiteness and objective indefiniteness}

Our thesis in this paper is that mathematics (including logic) can be used to
attack the problem of finding a realistic interpretation of (standard
Dirac-von-Neumann) quantum mechanics (QM). Mathematics itself contains a very
basic duality that can be associated with two meta-physical types of reality:

\begin{enumerate}
\item the common-sense notion of objectively definite reality assumed in
classical physics, and

\item the notion of objectively indefinite reality suggested by quantum physics.
\end{enumerate}

\noindent The problem of interpreting quantum mechanics \textit{is}
essentially the problem of making sense out of the notion of objective indefiniteness.

The approach taken here is to follow the lead of the mathematics of
partitions, first for sets (where things are relatively "clear and distinct")
and then for complex vector spaces where the mathematics of full QM resides.

There has long been the notion of subjective or epistemic indefiniteness
("cloud of ignorance") that is slowly cleared up with more discrimination and
distinctions (as in the game of Twenty Questions). But the vision of reality
that seems appropriate for quantum mechanics is \textit{objective or
ontological indefiniteness}. The notion of objective indefiniteness in QM has
been most emphasized by Abner Shimony (\cite{shim:reality},
\cite{shim:concept}, \cite{shim:worldview}).

\begin{quotation}
\noindent From these two basic ideas alone -- indefiniteness and the
superposition principle -- it should be clear already that quantum mechanics
conflicts sharply with common sense. If the quantum state of a system is a
complete description of the system, then a quantity that has an indefinite
value in that quantum state is objectively indefinite; its value is not merely
unknown by the scientist who seeks to describe the system. \cite[p.
47]{shim:reality}

\noindent The fact that in any pure quantum state there are physical
quantities that are not assigned sharp values will then mean that there is
\textit{objective indefiniteness} of these quantities. \cite[p. 27]%
{shim:worldview}
\end{quotation}

\noindent The view that a description of a superposition quantum state is a
\textit{complete} description means that the indefiniteness of a superposition
state is in some sense objective or ontological and not just subjective or epistemic.

In addition to Shimony's "objective indefiniteness" (the phrase also used by
Gregg Jaeger \cite{jaeger:objindef} and used here), other philosophers of
physics have suggested related ideas such as:

\begin{itemize}
\item Peter Mittelstaedt's "incompletely determined" quantum states with
"objective indeterminateness" \cite{mitt:kant},

\item Paul Busch and Gregg Jaeger's "unsharp quantum reality"
\cite{busch-jaeger:unsharp},

\item Paul Feyerabend's "inherent indefiniteness" \cite{feyerabend:micro},

\item Allen Stairs' "value indefiniteness" and "disjunctive facts"
\cite{stairs:disjfacts},

\item E. J. Lowe's "vague identity" and "indeterminacy" that is "ontic"
\cite{lowe:vagueid},

\item Steven French and Decio Krause's "ontic vagueness"
\cite{french:onticvagueness},

\item Paul Teller's "relational holism" \cite{teller:relholism}, and so forth.
\end{itemize}

Indeed, the idea that a quantum state is in some sense blurred or like a cloud
is now rather commonplace even in the popular literature. Thus the idea of
objective indefiniteness (described in various ways) is hardly new. Our goal
is to give the mathematical backstory to indefiniteness by showing how the
mathematical framework of QM can be built up or developed starting with the
(new) mathematical logic appropriate for indefiniteness that is
duality-related to the usual Boolean logic associated with classical definiteness.

\subsection{Mathematical description of indefiniteness = partitions}

How can indefiniteness be depicted mathematically? The basic idea is simple;
start with what is taken as full definiteness and then factor or quotient out
the surplus definiteness using an equivalence relation or partition.

Starting with some universe set $U$ of fully distinct and definite elements, a
partition $\pi=\left\{  B_{i}\right\}  $ (i.e., a set of disjoint blocks
$B_{i}$ that sum to $U$) collects together in a block (or cell) $B_{i}$ the
distinct elements $u\in U$ whose distinctness is to be ignored or factored
out, but the blocks are still distinct from each other. Each block represents
the elements that are the same in some respect (since each block is an
equivalence class in the associated equivalence relation on $U$), so the block
is indefinite between the elements within it. But different blocks are still
distinct from each other in that aspect.

\begin{example}
Consider the calculation of the binomial coefficient $\binom{N}{m}=\frac
{N!}{m!\left(  N-m\right)  !}$. The idea is to count the number of $m$-ary
subsets of an $N$-ary set ($m\leq N$) where the different orderings of the
otherwise same $m$-ary subset are surplus that need to be factored out. The
method of calculation is to first count the number of possible orderings of
the whole $N$-ary subset which is $N!=N\left(  N-1\right)  ...\left(
2\right)  \left(  1\right)  $. Then we want to quotient out the cases that are
distinct only because of different orderings. For any given ordering of the
$N$ elements, there are $m!$ ways to permute the first $m$ elements in the
given ordering--leaving the last $N-m$ elements the same. Thus we take the
first quotient by identifying any two of the $N!$ different orderings if they
differ only in a permutation of those first $m$ elements. Since there are $m!$
such permutations, there are now $N!/m!$ equivalence classes or blocks in the
resulting partition of the $N!$ orderings. But these equivalence classes still
count as distinct the different orderings of the last $N-m$ elements so we
further identify blocks which just have a permutation of the last $N-m$
elements to make larger blocks. Then the result is $\binom{N}{m}=\frac
{N!}{m!\left(  N-m\right)  !}$ blocks in the partition which is the number of
$m$-element subsets (which equals the number of $N-m$-element subsets) out of
an $N$-ary set disregarding the ordering of the elements.
\end{example}

In this example, the set of fully determinate alternatives are the $N!$
orderings of the $N$-element set. Then to consider the subsets of determinate
or definite cardinality $m$ (and thus the complementary subsets of definite
cardinality $N-m$), we must quotient out the number of possible orderings $m!$
and $\left(  N-m\right)  !$ to render the ordering of the elements in the
subsets indefinite or indeterminate.

\begin{example}
To be concrete, consider a set $\left\{  a,b,c,d\right\}  $ of $N=4$ elements
so the universe $U$ for fully distinct orderings has $4!=24$ elements
$\left\{  abcd,abdc,...\right\}  $. How many $2$-element subsets are there?
The first quotient groups together or identifies the orderings which only
permute the first $m=2$ elements so two of the blocks in that partition are
$\left\{  abcd,bacd\right\}  $ and $\left\{  abdc,badc\right\}  $, and there
are $N!/m!=24/2=12$ such blocks. Each block has the same final $N-m=2$
elements in the ordering so we further identify the blocks that differ only in
a permutation of those last $N-m$ elements. One of the blocks in that final
partition is $\left\{  abcd,bacd,abdc,badc\right\}  $ and there are $\frac
{N!}{m!\left(  N-m\right)  !}=\frac{24}{\left(  2\right)  \left(  2\right)
}=6$ such blocks with four elements in each block. Each block is distinct from
the other blocks in the first $m$ elements and in the last $N-m$ elements of
the orderings in the block so the block count is just the number of subsets of
$m$ elements (which equals the number of subsets of $N-m$ elements as well)
where each block is indefinite as to the ordering of elements within the first
$m$ elements and within the last $N-m$ elements.
\end{example}

Hermann Weyl makes the same point using an example slightly more complicated
than the binomial coefficient. He starts with a set or "aggregate $S$ ... of
elements each of which is in a definite state" \cite[p. 239]{weyl:phil} and
then considers a partition or equivalence relation whose $k$ blocks or classes
$C_{1},...,C_{k}$ can be thought of as boxes into which the $n$ elements of
$S$ may be placed (some boxes might be empty).

\begin{quotation}
\noindent A definite \textit{individual state} of the aggregate $S$ is then
given if it is known, for each of the $n$ marks $p$ [DE: which distinguish the
$n$ elements of $S$], to which of the $k$ classes [or boxes] the element
marked $p$ belongs. Thus there are $k^{n}$ possible \textit{individual states}
of $S$. If, however, no artificial differences between elements are introduced
by their labels $p$ and merely the intrinsic differences of state are made use
of, then the aggregate is completely characterized by assigning to each class
$C_{i}$ ($i=1,...,k$) the number $n_{i}$ of elements of $S$ that belong to
$C_{i}$. \cite[p. 239]{weyl:phil}
\end{quotation}

\noindent Those occupation numbers $n_{i}$ would then characterize "the
\textit{visible} or \textit{effective state} of the system $S$." \cite[p.
239]{weyl:phil} Thus Weyl points out that the mathematical treatment of
indefiniteness starts with the definiteness given here by the markings $p$ on
the $n$ elements distributed between the $k$ boxes $C_{i}$, and then one
erases the markings $p$ so the blocks or boxes in the partition have only an
occupation number $n_{i}$ with no distinctions between the $n_{i}$ elements in
each box $C_{i}$. When this scheme for representing indefiniteness is applied
in quantum mechanics, then it is an objective indefiniteness in that no
further differentiation between the elements of a box $C_{i}$ is possible.

\begin{quotation}
\noindent Since photons come into being and disappear, are emitted and
absorbed, they are individuals without identity. No specification beyond what
was previously called the effective state of the aggregate is therefore
possible. Hence the state of a photon gas is known when for each possible
state $\alpha$ of a photon the number $n_{\alpha}$ of photons in that state is
given (Bose-Einstein statistics of radiation). \cite[p. 246]{weyl:phil}
\end{quotation}

Thus within QM, the treatment of the indefiniteness due to the
indistinguishability of quantum particles of the same type is to artificially
treat them as distinct and then collect together or superpose the permutations
of the particles (in a totally symmetric or totally antisymmetric manner) that
factors out their supposed distinctness (see any QM text such as
\cite{cohen-tann:bothvol}).

Our point in this section is the general mathematical theme that
indefiniteness is described by taking a partition or quotient of a set of
definite entities. A partition is a mixture of indefiniteness and
definiteness. Each block is indefinite between the elements within it, but the
blocks of the partition are distinct from one another.

\subsection{Mathematical description of definiteness = subsets}

The common-sense classical view of reality is that it is completely definite
or determined and fully propertied all the way down. Every entity or thing
definitely has a property $P$ or definitely has the property $\lnot P$. Peter
Mittelstaedt quotes Immanuel Kant's treatment of the idea of complete determinateness:

\begin{quote}
Every thing as regards its possibility is likewise subject to the principle of
complete determination according to which if all possible predicates are taken
together with the contradictory opposites, then one of each pair of
contradictory opposites must belong to it. (Kant quoted in: \cite[p.
170]{mitt:kant})
\end{quote}

Given a universe set $U$, a predicate $P$ is represented by the subset
$S\subseteq U$ of elements that have the property, and the complement subset
$S^{c}=U-S$ represents the elements that have the property $\lnot P$.

\subsection{Two logics for the two types of reality}

The two mathematical concepts of subsets and partitions are thus associated
with two aforementioned metaphysical types of reality:

\begin{enumerate}
\item the common-sense notion of objectively definite reality assumed in
classical physics, and

\item the notion of objectively indefinite reality suggested by quantum mechanics.
\end{enumerate}

Subsets and quotient sets (or partitions) are mathematically \textit{dual}
concepts in the reverse-the-arrows sense of category-theoretic duality, e.g.,
a subset is the direct image of a set monomorphism while a set partition is
the inverse image of an epimorphism. This duality is familiar in abstract
algebra in the interplay of subobjects (e.g., subgroups, subrings, etc.) and
quotient objects. William Lawvere calls the general category-theoretic notion
of a subobject a \textit{part}, and then he notes: "The dual notion (obtained
by reversing the arrows) of `part' is the notion of partition."\cite[p.
85]{law:sets}

The logic appropriate for the usual notion of fully definite reality described
by subsets is the ordinary Boolean logic of subsets \cite{boole:lot} (usually
mis-specified as the special case of "propositional" logic). We have seen that
the other vision of objectively indefinite reality suggested by QM is
mathematically described by the equivalent concepts of quotient sets,
partitions, or equivalence relations. The Boolean logic of subsets has an
equally fundamental logic of the dual quotient sets, equivalence relations, or
partitions (\cite{ell:partitions} and \cite{ell:intropart}).\footnote{The
Boolean logic of subsets and the logic of partitions are equally fundamental
in that they are the two logics that take (subsets of and partitions on)
\textit{arbitrary} universe sets in their semantics. Other logics that have a
precise semantics, such as intuitionistic logic, have some additional
structures, such as topologies, order relations, accessibility relations, etc.
on the universe sets and thus those logics partake of the specific nature of
those additional structures.} The two logics of the dual concepts are
associated with the two visions of reality.

Boole developed not only the logic of subsets, he defined a normalized
counting measure on subsets of a finite universe set to yield the notion of
\textit{logical} finite probability theory. By making the same mathematical
moves in the logic of partitions, i.e., by defining a normalized counting
measure on the partitions on a finite universe set (where partitions are
represented as the complement of the associated equivalence relations), one
obtains a notion of \textit{logical} \textit{entropy} (\cite{ell:distinctions}
and \cite{ell:logicentrop}) that adds to the interpretation of QM
mathematics.\footnote{Although beyond the scope of this paper (see
\cite{ell:qmoversets}), logical information theory gives the interpretation of
the off-diagonal "coherences" in density matrices and the change in those
off-diagonal entries following a measurement as well as the mathematical
description of the non-unitary measurement process itself.}

Since our topic is to better understand objective indefiniteness, and thus to
better understand the reality described by QM, we will be following the math
of partitional concepts. The development of the mathematical concepts from the
logical level up to QM is based on the natural bridge between set concepts and
vector-space concepts using the notion of a basis set. We will transport
partitional concepts across that bridge in both directions. We will see that
the mathematics and logic of set partitions can be lifted or transported to
complex (inner product) vector spaces where it yields essentially the
mathematical machinery of QM (of course, not the specifically physical
postulates such as the Hamiltonian or the DeBroglie relationships). This shows
that the mathematical framework of QM is a type of \textit{logical} system
developed using complex vector spaces. The vector space concepts of full QM
can also be transported back to set-like vector spaces over $%
\mathbb{Z}
_{2}$ to yield a pedagogical model of "quantum mechanics over sets" or QM/sets
\cite{ell:qmoversets}. The probability calculus of QM/sets is a
non-commutative version of the usual Laplace-Boole logical finite probability
theory. The traffic in both directions supports the idea of interpreting QM in
terms of objective indefiniteness as illuminated by the logic and mathematics
of partitions \cite{ell:objindef}.

\subsection{Some imagery for objective indefiniteness}

In subset logic, each element of the universe set $U$ either definitely has or
does not have a given property $P$ (represented as a subset of the universe).
Moreover an element has properties all the way down so that two numerically
distinct entities must differ by some property as in Leibniz's principle of
the identity of indiscernibles.\cite{ladyman:indisc-qm} Change takes place by
the definite properties changing. For a hound to go from point $A$ to point
$B$, there must be some trajectory of definite ground locations from $A$ to
$B$.

In the logic of partitions, a partition $\pi=\left\{  B_{i}\right\}  $ is made
up of disjoint blocks $B_{i}$ whose union is the universe set $U$ (the blocks
are also thought of as the equivalence classes in the associated equivalence
relation). The blocks in a partition have been distinguished from each other,
but the elements within each block have not been distinguished from each other
by that partition, i.e., they are identified by the associated equivalence
relation. Hence each block can be viewed as the set-theoretic version of a
superposition of the distinct elements in the block. When more distinctions
are made (the set-version of a measurement), the blocks get smaller and the
partitions (set-version of mixed states) become more refined until the
discrete partition $\mathbf{1}=\left\{  \left\{  u\right\}  :\left\{
u\right\}  \subseteq U\right\}  $ is reached where each block is a singleton
(the set-version of a non-degenerate measurement yielding a completely
decoherent mixed state). Change takes place by some attributes becoming more
definite and other (incompatible) attributes becoming less definite. For a
hawk, as opposed to a hound, to go from point $A$ to point $B$, it would go
from a definite perch at $A$ into a flight of indefinite ground locations, and
then would have a definite perch again at $B$.\footnote{The "flights and
perchings" metaphor is from William James \cite[p. 158]{james:psy} and
according to Max Jammer, that description "was one of the major factors which
influenced, wittingly or unwittingly, Bohr's formation of new conceptions in
physics." \cite[p. 178]{jammer:concept} The hawks and hounds pairing comes
from Shakespeare's Sonnet 91.}%

\begin{center}
\includegraphics[
height=2.5122in,
width=5.506in
]%
{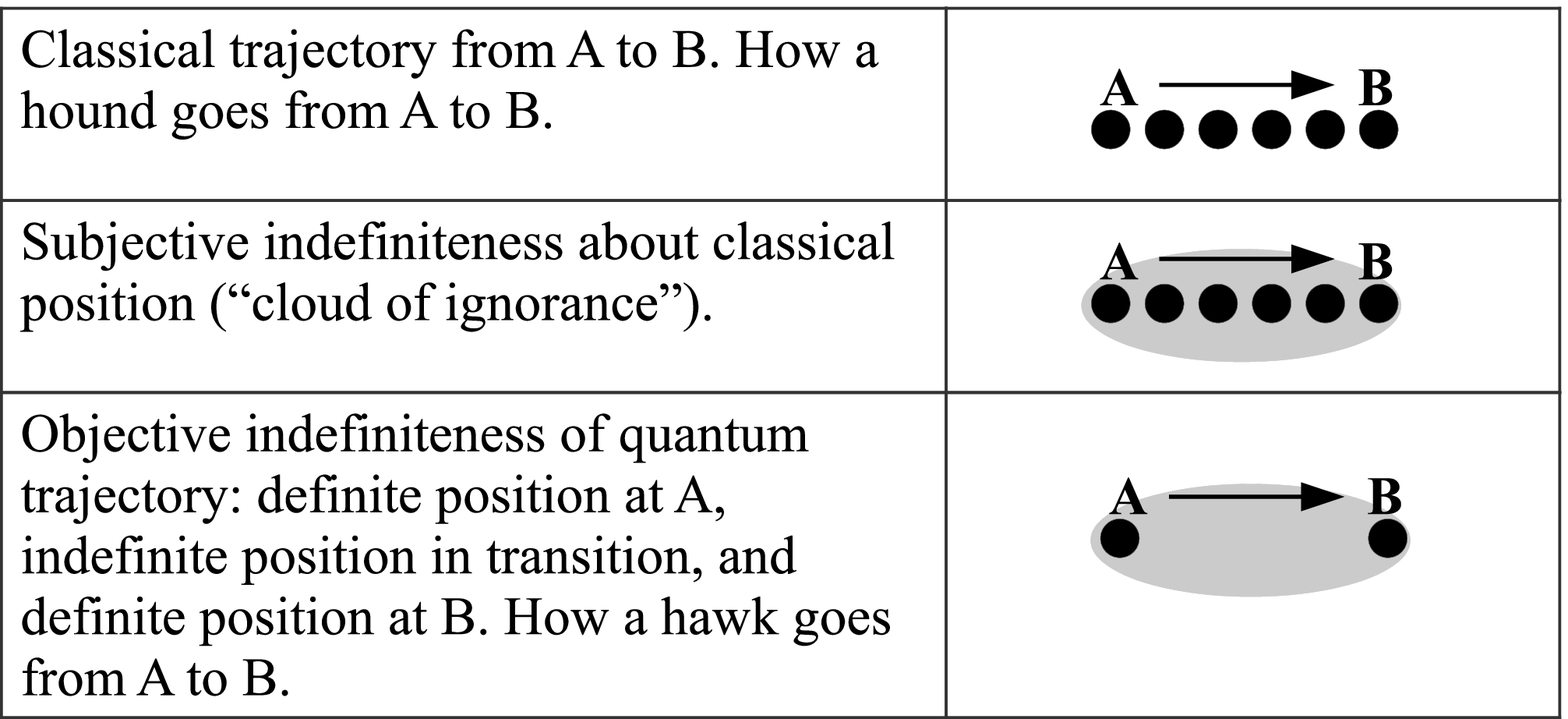}%
\\
Figure 1: How a hound and a hawk go from $A$ to $B$%
\end{center}

The imagery of having a sharp focus versus being out-of-focus could also be
used if one is clear that it is the reality itself that is in-focus or
out-of-focus, not just the image through, say, a microscope. A classical
trajectory is like a moving picture of sharp or definite in-focus realities,
whereas the quantum trajectory starts with a sharply focused reality, goes out
of focus, and then returns to an in-focus reality (by a measurement).

In the objective indefiniteness interpretation, a subset $S\subseteq U$ of a
universe set $U$ should be thought of as a single indefinite element $S$ that
is only \textit{represented} as a subset of fully definite elements $\left\{
u:u\in S\right\}  $--just as a single superposition vector is represented as a
weighted vector sum of certain basis of eigen ("eigen" should be translated as
"definite" here) vectors. Abner Shimony (\cite{shim:reality} and
\cite{shim:concept}), in his description of a superposition state as being
objectively indefinite, sometimes used Heisenberg's \cite{heisen:phy-phil}
language of "potentiality" and "actuality" to describe the relationship of the
eigenvectors that are superposed to give an objectively indefinite
superposition. This terminology could be adapted to the case of the sets. The
singletons $\left\{  u\right\}  \subseteq S$ are "potential" in the
objectively indefinite superposition$S$, and, with further distinctions, the
indefinite element $S$ might "actualize" to $\left\{  u\right\}  $ for one of
the "potential" $\left\{  u\right\}  \subseteq S$. Starting with $S$, the
other $\left\{  u\right\}  \nsubseteqq S$ (i.e., $u\notin S$) are not
"potentialities" that could be "actualized" with further distinctions.

This terminology is, however, somewhat misleading since the indefinite element
$S$ is perfectly actual(in the objectively indefinite interpretation); it is
only the multiple eigen-elements $\left\{  u\right\}  \subseteq S$ that are
"potential" until "actualized" by some further distinctions. A non-degenerate
measurement is not a process of a potential entity becoming an actual entity,
it is a process of an actual indefinite element becomes an actual definite
element. Since a distinction-creating measurement goes from actual indefinite
to actual definite, the potential-to-actual language of Heisenberg should only
be used with proper care--if at all.

Note that there are two conceptually distinct connotations for the
mathematical subset $S\subseteq U$. In the classical interpretation, it is a
set of fully definite elements of $u\in S$. In the quantum interpretation of a
subset $S$, it is a single indefinite element that with further distinctions
could become one of the singleton definite-elements or eigen-elements
$\left\{  u\right\}  \subseteq S$.

Consider a three-element universe $U=\left\{  a,b,c\right\}  $ and a partition
$\pi=\left\{  \left\{  a\right\}  ,\left\{  b,c\right\}  \right\}  $. The
block $S=\left\{  b,c\right\}  $ is objectively indefinite between $\left\{
b\right\}  $ and $\left\{  c\right\}  $ so those singletons are its
"potentialities" in the sense that a distinction could result in either
$\left\{  b\right\}  $ or $\left\{  c\right\}  $ being "actualized." However
$\left\{  a\right\}  $ is not a "potentiality" when one is starting with the
indefinite element $\left\{  b,c\right\}  $.

Note that this objective indefiniteness of $\left\{  b,c\right\}  $ is
\textit{not} well-described as saying that indefinite pre-distinction element
is "simultaneously both $b$ and $c$" (like the common misdescription of the
undetected particle "going through both slits" in the double-slit experiment);
instead it is indefinite between $b$ and $c$. That is, a superposition of two
sharp eigen-alternatives should \textit{not} be thought of like a
double-exposure photograph which has two fully definite images (e.g.,
simultaneously a picture of say $b$ and $c$). Instead of a double-exposure
photograph, the superposition should be thought of as representing a blurred
or incomplete reality that with further distinctions could sharpen to either
of the sharp realities. But there must be some way to indicate which sharp
realities could be obtained by making further distinctions (measurements), and
that is why the blurred or cloud-like indefinite reality is represented by
mathematically superposing the sharp potentialities.

This point might be illustrated using some Guy Fawkes masks.%

\begin{center}
\includegraphics[
height=2.9066in,
width=3.164in
]%
{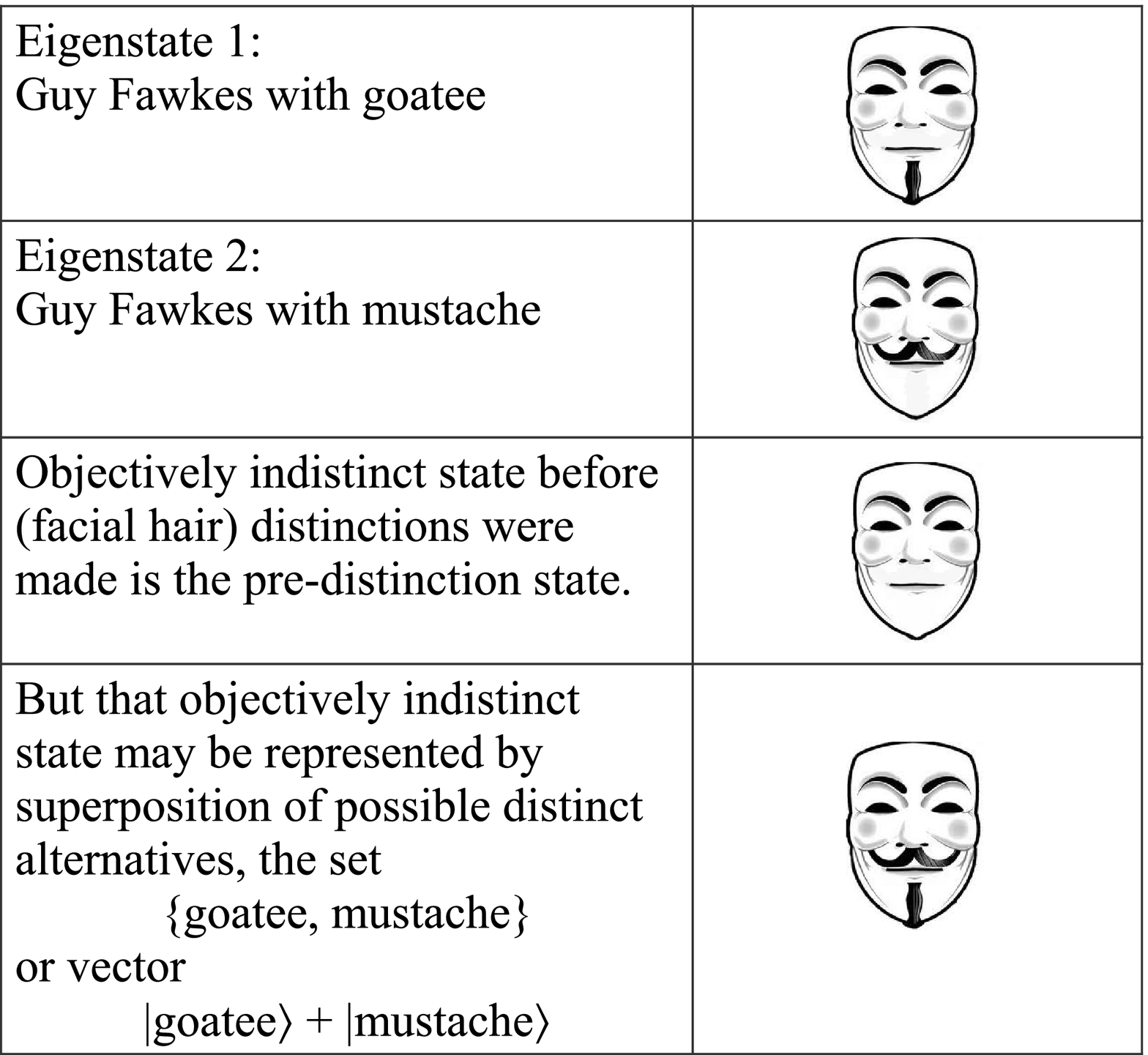}%
\end{center}

\begin{center}
Figure 2: Objectively indefinite pure state \textit{represented} as
superposition of distinct eigen-alternatives
\end{center}

Instead of a double-exposure photograph, a superposition representation might
be thought of as "a photograph of clouds or patches of fog." (Schr\"{o}dinger
quoted in: \cite[p. 66]{fine:shaky}) Schr\"{o}dinger distinguishes a
"photograph of clouds" from a blurry photograph presumably because the latter
might imply that it was only the photograph that was blurry while the
underlying objective reality was sharp. The "photograph of clouds" imagery for
a superposition connotes a clear and complete photograph of an objectively
"cloudy" or indefinite reality. Regardless of the (imperfect) imagery, one
needs some way to indicate what are the definite eigen-elements that could be
"actualized" from a single indefinite element $S$, and \textit{that} is the
role in the set case of conceptualizing a subset $S$ as a collecting together
or "superposing" certain "potential" eigen-states, i.e., the singletons
$\left\{  u\right\}  \subseteq S$.

\subsection{The two lattices}

The two dual subset and partition logics are modeled by the two lattices (or,
with more operations, algebras) of subsets and of partitions.\footnote{The two
logics and lattices are "dual" or "duality-related" in the sense of the
subset-partition duality.} The conceptual duality between the lattice of
subsets (the lattice part of the Boolean algebra of subsets of $U$) and the
lattice of partitions could be described (again following Heisenberg) using
the rather meta-physical notions of \textit{substance}\footnote{Heisenberg
identifies "substance" with energy.
\par
\begin{quotation}
\noindent Energy is in fact the substance from which all elementary particles,
all atoms and therefore all things are made, and energy is that which moves.
Energy is a substance, since its total amount does not change, and the
elementary particles can actually be made from this substance as is seen in
many experiments on the creation of elementary particles. \cite[p.
63]{heisen:phy-phil}
\end{quotation}
} and \textit{form }(as in in-form-ation)--which might be compared to the
terms "matter" or "objects" on one hand and "structure" on the other hand in
some modern metaphysical discussions.\footnote{See McKenzie
\cite{mckenzie:priority} and the references therein to ontic structural
realism.}

For each lattice where $U=\left\{  a,b,c\right\}  $, start at the bottom and
move towards the top.%

\begin{center}
\includegraphics[
height=2.1669in,
width=5.5824in
]%
{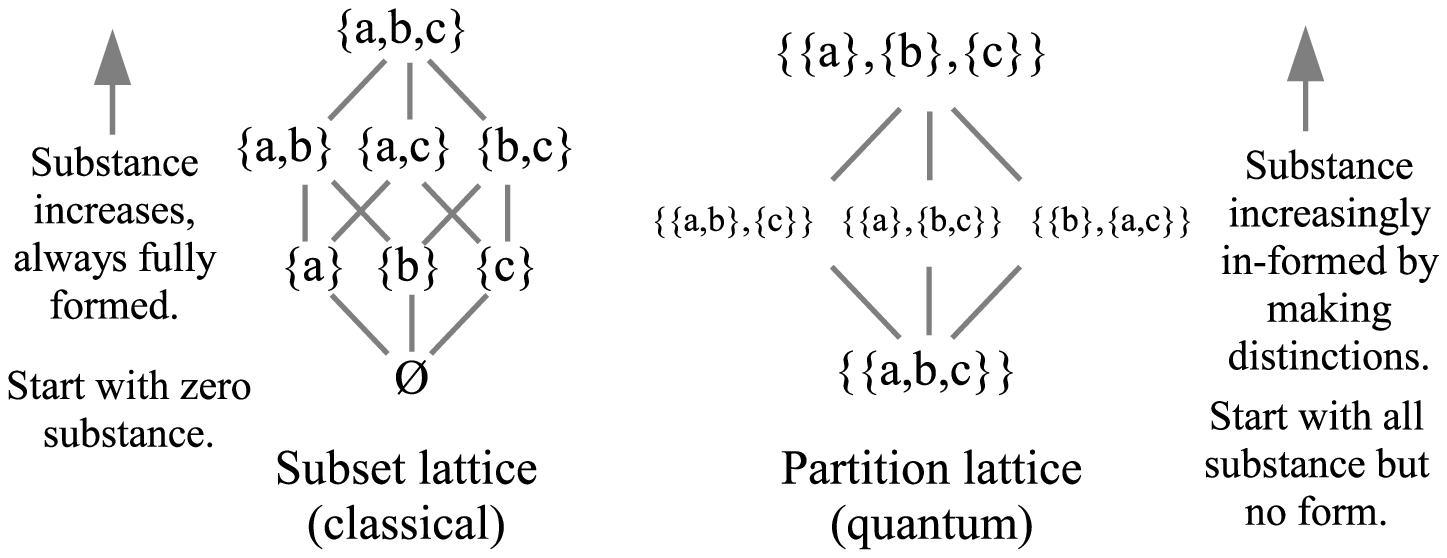}%
\\
Figure 3: Conceptual duality between the subset and partition logics
\end{center}

At the bottom of the Boolean lattice is the empty set $\emptyset$ which
represents no substance. As one moves up the lattice, new elements of
substance always with fully definite properties are created until finally one
reaches the top, the universe $U$. Thus new substance is created in moving up
the lattice but each element is fully formed and distinguished in terms of its properties.

At the bottom of the partition lattice is the indiscrete partition or "blob"
$\mathbf{0=}\left\{  U\right\}  $ (where the universe set $U$ makes one block)
which represents all the substance but with no distinctions to in-form the
substance.\footnote{The "blob" is the set-version of a pure state in QM prior
to a distinctions-creating measurement that decoheres the pure state into
non-blob partitions analogous to a mixed state (see \cite{ell:qmoversets} for
spelling this out using density matrices).} As one moves up the lattice, no
new substance is created but distinctions objectively in-form the indistinct
elements as they become more and more distinct, until one finally reaches the
top, the discrete partition $\mathbf{1}$, where all the eigen-elements of $U$
have been fully distinguished from each other.\footnote{This notion of logical
in-formation as distinctions is based on partition logic just as logical
probability is based on subset logic (\cite{ell:distinctions} and
\cite{ell:logicentrop}). That is, the \textit{logical entropy} of a partition
is the normalized counting measure of the distinctions of a partition
(represented as a binary relation) just as the Laplace-Boole \textit{logical
probability} of a subset is the normalized counting measure on the subsets
(events) of the finite universe set (set of equiprobable outcomes).} It was
previously noted that a partition combines indefiniteness (within blocks) and
definiteness (between blocks). At the top of the partition lattice, the
discrete partition $\mathbf{1}=\left\{  \left\{  u\right\}  :\left\{
u\right\}  \subseteq U\right\}  $ is the result making all the distinctions to
eliminate the indefiniteness. Thus one ends up at the "same" place
(macro-universe of distinguished elements) either way, but by two totally
different but dual ways.\footnote{In treating the universe $U=\left\{
u,u^{\prime},...\right\}  $ and the discrete partition $\mathbf{1}=\left\{
\left\{  u\right\}  ,\left\{  u^{\prime}\right\}  ,...\right\}  $ as the
"same" we are neglecting the distinction between $u$ and $\left\{  u\right\}
$ for $u\in U$.}

The progress from bottom to top of the two lattices could also be described as
two creation stories.

\begin{itemize}
\item \noindent\textit{Subset creation story}: \textquotedblleft In the
Beginning was the Void\textquotedblright, and then elements are created, fully
propertied and distinguished from one another, until finally reaching all the
elements of the universe set $U$.

\item \noindent\textit{Partition creation story}: \textquotedblleft In the
Beginning was the Blob\textquotedblright, which is an undifferentiated
\textquotedblleft substance,\textquotedblright\ and then there is a "Big Bang"
where elements (\textquotedblleft its\textquotedblright) are created by the
substance being objectively in-formed (objectified information) by the making
of distinctions (e.g., breaking symmetries) until the result is finally the
singletons which designate the elements of the universe $U$.
\end{itemize}

These two creation stories might also be illustrated as follows.%

\begin{center}
\includegraphics[
height=1.2652in,
width=5.3059in
]%
{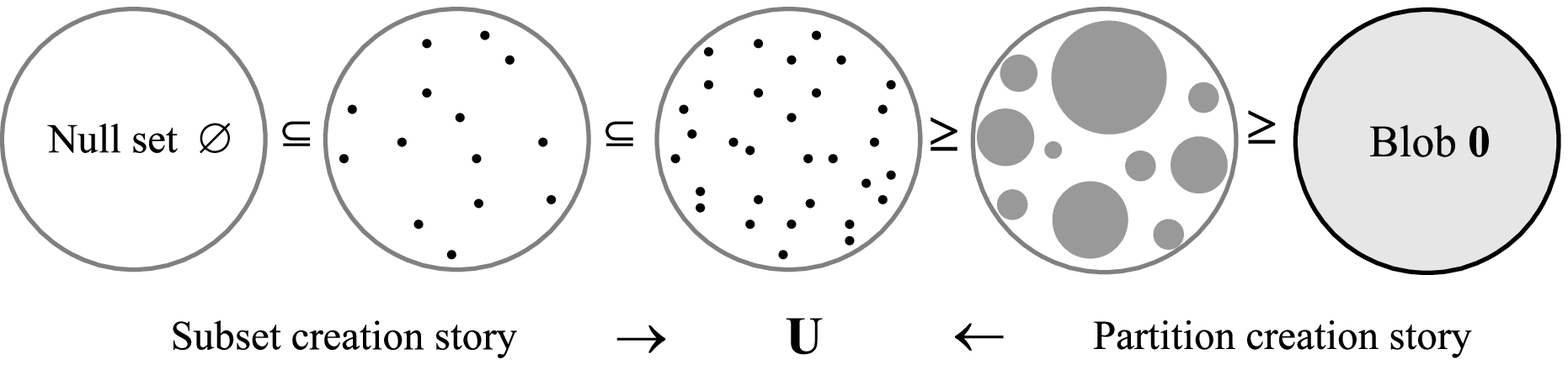}%
\end{center}

\begin{center}
Figure 4: Two creation stories
\end{center}

One might think of the universe $U$ (in the middle of the above picture) as
the macroscopic world of definite entities that we ordinarily experience.
Common sense and classical physics assumes, as it were, the subset creation
story on the left. But \textit{a priori,} it could just as well have been the
dual story, the partition creation story pictured on the right, that leads to
the \textit{same} macro-picture $U$.

Since partitions are the mathematical expression of indefiniteness, our
strategy is to first show where set partitions come from and then to "lift" or
"transport" the partitional machinery to vector spaces using the notion of a
basis set of a vector space. The result is essentially the mathematical
machinery of quantum mechanics--all of which shows how quantum mechanics can
be interpreted using the objective indefiniteness conception of reality that
is associated at the logical level with partition logic.

\section{Whence set partitions?}

\subsection{Set partitions from set attributes}

Take the universe set as some specific set of people, say, in a room. People
have numerical attributes like weight, height, or age as well as non-numerical
attributes with other values such place of birth, family name, and country of
citizenship. Abstractly an \textit{attribute on a universe set} $U$ is a
function $f:U\rightarrow R$ from $U$ to some set of values $R$ (usually the
reals $%
\mathbb{R}
$). In subset logic, an element $u\in U$ either has a \textit{property}
represented by a subset $S\subseteq U$ or not; in partition logic, an
attribute $f$ assigns a value $f\left(  u\right)  $ to each $\left\{
u\right\}  \subseteq U$. The two concepts of a property and an attribute
overlap for binary attributes where the attribute might be represented by the
characteristic function $\chi_{S}:U\rightarrow2=\left\{  0,1\right\}  $ of a
subset $S\subseteq U$.\footnote{To be technically precise, a subset $S$ is
given by a binary attribute $\chi_{S}:U\rightarrow2=\left\{  0,1\right\}  $
\textit{plus} the designation of an element $1\in2$ so that $S=\chi_{S}%
^{-1}\left(  1\right)  $ as in Lawvere's well-known subobject-classifier
diagram \cite[p. 39]{law:sets}.}

Each attribute $f:U\rightarrow R$ on a universe $U$ determines the
\textit{inverse-image partition} $f^{-1}=\left\{  f^{-1}\left(  r\right)
\neq\emptyset:r\in R\right\}  $. Attributes are one way to define a partition
on a set $U$. Since this method of defining a partition starts with a
numerical attribute $f\left(  u\right)  $ already assigned to each $u\in U$,
it may be called the \textit{top-down} method.

\subsection{Set partitions from set representations of groups}

Another more "bottom-up" way to define a partition on $U$ is to map the
elements $u\in U$ to "similar" (i.e., same block) elements $u^{\prime}$ by
some set of transformations $G=\left\{  t:U\rightarrow U\right\}  $. This
defines a binary relation: $uGu^{\prime}$ if there exists a $t\in G$ such that
$t\left(  u\right)  =u^{\prime}.$ In order to define a partition, the binary
relation $uGu^{\prime}$ has to be an equivalence relation so the blocks of the
partition are the equivalence classes. The three requirements for an
equivalence relation are reflexivity, symmetry, and transitivity.

\begin{itemize}
\item For the relation to be reflexive, i.e., $uGu$ for all $u\in U$, it is
sufficient for the set of transformations $G$ to contain the identity
transformation $1_{U}:U\rightarrow U$.

\item For the relation to be symmetric, i.e., $uGu^{\prime}$ implies
$u^{\prime}Gu$, it is sufficient for each $t\in G$ to have an inverse
$t^{-1}\in G$ where $U\overset{t}{\longrightarrow}U\overset{t^{-1}%
}{\longrightarrow}U=1_{U}=U\overset{t^{-1}}{\longrightarrow}%
U\overset{t}{\longrightarrow}U$.

\item For the relation to be transitive, i.e., $uGu^{\prime}$ and $u^{\prime
}Gu^{\prime\prime}$ imply $uGu^{\prime\prime}$, it is sufficient for each
$t,t^{\prime}\in G$ that $t^{\prime}t:U\overset{t}{\longrightarrow
}U\overset{t^{\prime}}{\longrightarrow}U$ is also in $G$.
\end{itemize}

These three conditions, the existence of the identity, the existence of an
inverse, and closure under composition, define a \textit{transformation group}
$G=\left\{  t:U\rightarrow U\right\}  $, i.e., a \textit{group action} on a
set $U$. Equivalently, a \textit{set representation} of a group $G$ is given
by a group homomorphism $T:G\rightarrow S\left(  U\right)  $, where $S\left(
U\right)  $ is the symmetric group of permutations $t$ of the set $U$ (and
where the transformation group $\left\{  t:U\rightarrow U\right\}  \subseteq
S\left(  U\right)  $ is the image of the map). An \textit{abstract group}
satisfies these three conditions where the composition is also required to be
associative in the sense that for any $t,t^{\prime},t^{\prime\prime}\in G$,
$\left(  t^{\prime\prime}t^{\prime}\right)  t=t^{\prime\prime}\left(
t^{\prime}t\right)  $. For a transformation group, the composition is
automatically associative.

This connection between groups and equivalence relations or partitions is
well-known, e.g., \cite{castellani:symeq}, and is probably as old as the
notion of a group. Instead of elements $u,u^{\prime}\in U$ being collected in
the same block by have the same attribute value $f\left(  u\right)  =f\left(
u^{\prime}\right)  $, the group transformations take any element $u$ to a
"similar" or "symmetric" element $t\left(  u\right)  =u^{\prime}$. A subset
$S\subseteq U$ is \textit{invariant under }$G$ if for any $t\in G$, $t\left(
S\right)  \subseteq S$. A minimal invariant subset is an \textit{orbit}, and
the partition defined by the transformation group $G$ is the \textit{set
partition of orbits}. That is the\textit{ bottom-up} method of defining a set
partition since we don't begin with some attribute-value already assigned to
the elements of $U$.

What is the significance of the blocks in the partition of minimal invariant
subsets? Often the treatment of symmetry groups focuses on what is invariant
or conserved, e.g., the perspective of Noether's theorem \cite{brad:symm}.

There is \textit{another} perspective with which to view the representations
of symmetry groups. To represent an indefinite reality, there is first some
notion of the fully definite eigen-alternatives that are then collected
together or superposed to represent something indefinite between those alternatives.

\begin{center}
\textit{What determines the set of definite eigen-alternatives}?
\end{center}

\noindent Given a set of symmetries on a set, in what different ways can there
be distinct subsets that still satisfy the constraints of the symmetry
operations? The minimal invariant subsets or orbits of a set representation of
a symmetry group provide the answer to that question about the variety of
"atomic" eigen-forms consistent with the symmetries.

This question and the answer become more significant when we move beyond
structure-less sets to linear vector spaces. The minimal invariant subsets,
the orbits, then become the minimal invariant subspaces, the irreducible
subspaces, which are the carriers of the irreducible representations or irreps
in vector space representations of groups.

\subsection{Set partitions from other set partitions}

The notion of distinguishability (in principle, making a distinction) has long
been recognized as fundamental in QM.

\begin{quotation}
\noindent If you could, in principle, distinguish the alternative final states
(even though you do not bother to do so), the total, final probability is
obtained by calculating the probability for each state (not the amplitude) and
then adding them together. If you cannot distinguish the final states even in
principle, then the probability amplitudes must be summed before taking the
absolute square to find the actual probability. \cite[3-9]{feynman:vol-III}
\end{quotation}

In the foregoing, we have frequently referred to the making of
\textit{distinctions} as the set version of a
measurement.\footnote{Technically, a "distinction" of a partition
$\pi=\left\{  B\right\}  $ on $U$ is an ordered pair $\left(  u,u^{\prime
}\right)  $ of elements of $U$ in different blocks of the partition. The set
of distinctions, $\operatorname*{dit}\left(  \pi\right)  $, of a partition
$\pi$ is called a \textit{partition relation} (or \textit{apartness relation
}in computer science) and is just the complement of the associated equivalence
relation. The notion of a distinction of a partition is the partition logic
analogue of an element of a subset in subset logic. For instance, given two
partitions $\pi=\left\{  B\right\}  $ and $\sigma=\left\{  C\right\}  $ on a
universe set and two subsets $S$ and $T$ of a universe set, the partition join
$\pi\vee\sigma$ combines the \textit{distinctions} of the partitions, i.e.,
$\operatorname*{dit}\left(  \pi\vee\sigma\right)  =\operatorname*{dit}\left(
\pi\right)  \cup\operatorname*{dit}\left(  \sigma\right)  $, just as the
subset join or union $S\cup T$ combines the \textit{elements} of the subsets
(see \cite{ell:partitions} or \cite{ell:intropart} for further developments).}
What is the mathematical operation for making the distinctions (as in a
measurement)? It is the\textit{ join operation} from partition logic. But
before two set partitions can be joined to form a more refined partition with
more distinctions, they must be \textit{compatible} in the sense of being
defined on the same universe set. If two set partitions $\pi=\left\{
B\right\}  $ and $\sigma=\left\{  C\right\}  $ are compatible, i.e., are
partitions of the same universe $U$, then their \textit{join} $\pi\vee\sigma$
is the set partition whose blocks are the non-empty intersections $B\cap C$.

Since two set attributes $f:U\rightarrow%
\mathbb{R}
$ and $g:U^{\prime}\rightarrow%
\mathbb{R}
$ define two inverse image partitions $\left\{  f^{-1}\left(  r\right)
\right\}  $ and $\left\{  g^{-1}\left(  s\right)  \right\}  $ on their
domains, we need to extend the concept of compatible partitions to the
attributes that define the partitions. That is, two attributes $f:U\rightarrow%
\mathbb{R}
$ and $g:U^{\prime}\rightarrow%
\mathbb{R}
$ are \textit{compatible} if they have the same domain $U=U^{\prime}$.

Given two compatible set attributes $f:U\rightarrow%
\mathbb{R}
$ and $g:U\rightarrow%
\mathbb{R}
$, the join of their eigenspace partitions has as blocks the non-empty
intersections $f^{-1}\left(  r\right)  \cap g^{-1}\left(  s\right)  $. Each
block in the join of the eigenspace partitions could be characterized by the
ordered pair of eigenvalues$\left(  r,s\right)  $. An eigenvector of both $f$,
$S\subseteq f^{-1}\left(  r\right)  $, and of $g$, $S\subseteq g^{-1}\left(
s\right)  $, would be a simultaneous eigenvector: $S\subseteq f^{-1}\left(
r\right)  \cap g^{-1}\left(  s\right)  $.

A set of compatible set attributes is said to be \textit{complete} if the join
of their partitions is discrete, i.e., the blocks have cardinality $1$. A
\textit{Complete Set of Compatible Attributes} or CSCA characterizes the
singletons $\left\{  u\right\}  \subseteq U$ by the ordered $n$-tuple $\left(
r,...,s\right)  $ of attribute values.

All this machinery of set partitions can be lifted or transported to vector
spaces to give the mathematical machinery of QM.\footnote{In QM, the extension
of concepts on finite dimensional Hilbert space to infinite dimensional ones
is well-known. Since our expository purpose is conceptual rather than
mathematical, we will stick to finite dimensional spaces.}

\section{Lifting partition concepts from sets to vector spaces}

\subsection{The basis principle}

There is a natural part-of-the-folklore bridge or ladder connecting set
concepts to vector-space concepts. The basic idea is that a vector $v=\sum
_{i}\alpha_{i}b_{i}$, represented in terms of a set $\left\{  b_{i}\right\}  $
of basis vectors, is a $K$\textit{-valued set} where each element $b_{i}$ in
the basis set takes a value $c_{i}$ in the base field $K$. Given a set
concept, the \textit{basis principle} is that one can generate the
corresponding vector-space concept by applying the set concept to a basis set
and seeing what it generates. Starting with the set concept of cardinality,
one arrives at the corresponding vector-space concept by applying the set
concept to a basis set to arrive at the cardinality of the basis set. After
checking that all bases have the same cardinality, this yields the
vector-space notion of \textit{dimension}. Thus the cardinality of a set lifts
not to the cardinality of a vector space but to its dimension.

Some of the lifting is accomplished by the free vector space functor from the
category of sets to the category of vector spaces over a given field $K$. A
set $U$ is carried by this functor to the vector space $K^{U}$ spanned by the
Kronecker delta basis $\left\{  \delta_{u}:U\rightarrow K\right\}  _{u\in U}$
where $\delta_{u}\left(  u^{\prime}\right)  =0$ for $u^{\prime}\neq u$ and
$\delta_{u}\left(  u\right)  =1$. A set $U$ of a certain cardinality thus
generates a vector space $K^{U}$ of the same dimension.

\subsection{What is a vector space partition?}

A partition $\pi=\left\{  B\right\}  $ on a set $U$ is a set of subsets whose
direct sum (i.e., disjoint union) is the whole set, i.e., a direct sum
decomposition of the set. The corresponding vector space concept is a set of
subspaces of a vector space whose direct sum is the vector space, i.e., a
\textit{direct sum decomposition} of the vector space. In terms of the basis
principle, we could apply the set partition $\pi=\left\{  B\right\}  $ of a
set $U$ to a basis set $\left\{  b_{u}\right\}  _{u\in U}$, then each block
$B$ generates a subspace $V_{B}$ and the set of subspaces $\left\{
V_{B}\right\}  _{B\in\pi}$ is a direct sum decomposition of the vector space
spanned by the basis set. Thus the lift or transport of the concept of a set
partition is a direct sum decomposition of a vector space. In particular, it
is \textit{not} a set partition of a vector space that is compatible with the
vector space operations, i.e., a quotient space $V/W$ as would be defined by
each subspace $W\subseteq V$ with the equivalence relation $v\thicksim
v^{\prime}$ if $v-v^{\prime}\in W$. While a partition on a set is essentially
the same as a quotient set (or equivalence relation on the set), the
vector-space lift of a set partition is not a quotient vector space but a
direct sum decomposition of a vector space. In lifting or transporting the
partitional concepts for sets to vector spaces, we are making the choices
guided by the set-to-basis-set connection which yields the mathematical
machinery of quantum mechanics.

This is not particularly new; it is part of the mathematical folklore. Hermann
Weyl outlined the lifting program by first considering an attribute on a set,
which defined the set partition or "grating" \cite[p. 255]{weyl:phil} of
elements with the same attribute-value. Then he moved to the quantum case
where the set or "aggregate of $n$ states has to be replaced by an
$n$-dimensional Euclidean vector space" \cite[p. 256]{weyl:phil}%
.\footnote{Note the lift from sets to vector spaces using the basis principle
where the cardinality $n$ becomes dimension $n$.} The appropriate notion of a
partition or "grating" is a "splitting of the total vector space into mutually
orthogonal subspaces" so that "each vector $\overrightarrow{x}$ splits into
$r$ component vectors lying in the several subspaces" \cite[p. 256]%
{weyl:phil}, i.e., a direct sum decomposition of the space, where the
subspaces are the eigenspaces of an observable operator.

Weyl's grating metaphor also lends itself to (our own example of) seeing
measurement of the, say, 'regular polygonal shape' of an 'indefinite blob of
dough' as it randomly falls through a opening in a grating to take on that
'polygonal shape' (with the attribute-value or eigenvalue being the number of
regular sides $\lambda=3,4,5,6$).%

\begin{center}
\includegraphics[
height=2.0946in,
width=2.4085in
]%
{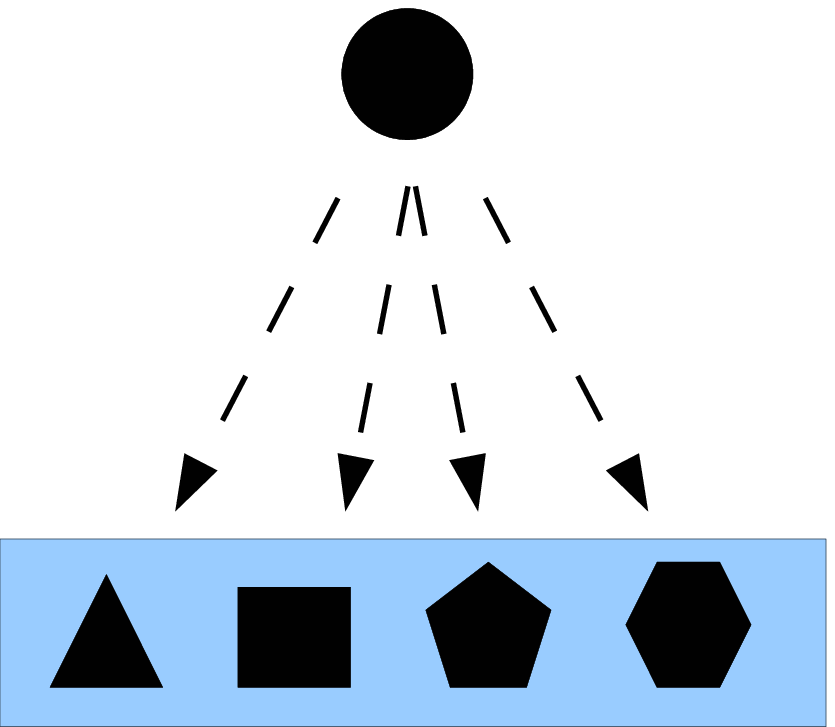}%
\end{center}

\begin{center}
Figure 5: Imagery of measurement as randomly giving an indefinite blob of
dough a definite eigen-shape.
\end{center}

\noindent Note how the blob of dough is objectively indefinite between the
regular polygonal shapes and does not simultaneously have all those shapes
even though it might be mathematically represented as the set $\left\{
\blacktriangle,\blacksquare,\ldots\right\}  $ or the superposition vector
$\blacktriangle+\blacksquare+\ldots$ in a certain space.

\subsection{What is a vector space attribute?}

A set attribute is a function $f:U\rightarrow%
\mathbb{R}
$ (where the set of values is taken as the reals). The inverse-image
$f^{-1}\left(  r\right)  \subseteq U$ of each value $f(u)=r$ is a subset where
the attribute has the same value, and those subsets form a set partition.
Given a basis set $\left\{  b_{u}\right\}  _{u\in U}$ of a vector space $V$
over a field $K$, we can apply a set attribute $f:\left\{  b_{u}\right\}
_{u\in U}\rightarrow K$ to the basis set and see what it generates. One
possibility is to linearly extend the function $f^{\ast}\left(  b_{u}\right)
=f\left(  b_{u}\right)  $ to the whole space to obtain a linear functional
$f^{\ast}:V\rightarrow K$. But a linear functional defines a quotient space
$V/f^{\ast-1}\left(  0\right)  $, not a vector space partition.

The same information $f:\left\{  b_{u}\right\}  _{u\in U}\rightarrow K$
\textit{also} defines $\hat{f}\left(  b_{u}\right)  =f\left(  b_{u}\right)
b_{u}$ which linearly extends to a \textit{linear operator} $\hat
{f}:V\rightarrow V$. The given basis vectors $\left\{  b_{u}\right\}  $ are
eigenvectors of the operator $\hat{f}$ with the eigenvalues $f\left(
b_{u}\right)  $, and the eigenspaces are the subspaces where the operator has
the same eigenvalue. The eigenvectors span the whole space so we see that the
lift or transport of a set attribute, which defines a set partition, is a
vector space linear operator whose eigenspaces are a vector space partition
(i.e., direct sum decomposition) of the whole space, i.e., a
\textit{diagonalizable linear operator}.\footnote{A diagonalizable linear
operator is the lift of a set attribute $f$ that is a total function, so a
non-diagonalizable linear operator is the lift of a partial function $f$.}

\section{Whence vector-space partitions?}

\subsection{Vector-space partitions from vector-space attributes}

Given a diagonalizable linear operator $L:V\rightarrow V$, where $V$ is a
finite-dimensional vector space over a field $K$ and where $\lambda
_{1},...,\lambda_{k}$ are the distinct eigenvalues, then there are projection
operators $P_{i}$ for $i=1,...,k$ such that:

\begin{enumerate}
\item $L=\sum_{i=1}^{k}\lambda_{i}P_{i}$;

\item $I=\sum_{i=1}^{k}P_{i}$;

\item $P_{i}P_{j}=0$ for $i\neq j$; and

\item the range of $P_{i}$ is the eigenspace $V_{i}$ for the eigenvalue
$\lambda_{i}$ for $i=1,...,k$. \cite[Theorem 8, p. 172]{hk:la}
\end{enumerate}

What is the vector space partition \textit{canonically} defined by a
diagonalizable linear operator? Any basis of eigenvectors could be seen as
defining a direct sum of the one-dimensional subspaces spanned by those
eigenvectors. But those subspaces are far from unique. But if we group
together all the eigenvectors with the same eigenvalue (i.e., use the top-down
method to define a vector-space partition), then they span the eigenspaces. It
is the set of eigenspaces $\left\{  V_{i}\right\}  $ that gives the unique
canonical direct-sum decomposition or vector-space partition defined by a
(diagonalizable) linear operator. This standard linear algebra result holds
for any base field, but for QM, the base field is the complex numbers $%
\mathbb{C}
$. In order for the eigenvalues to always be real, the diagonalizable linear
operators are required to be \textit{Hermitian} (or \textit{self-adjoint},
i.e., equal to their conjugate transposes).

\subsection{Vector-space partitions from vector-space representations of
groups}

A \textit{vector-space representation} of an abstract group $G$ is a group
homomorphism $T:G\rightarrow GL(V)$ where $GL\left(  V\right)  $ is the group
of invertible linear transformations $V\rightarrow V$ of a vector space $V$
over the complex numbers. Here again, the idea is to define a (vector-space)
partition by a (linear) group of transformations $T_{g}:V\rightarrow V$ that
map elements $v\in V$ to similar or symmetric elements $T_{g}\left(  v\right)
$. A subspace $W\subseteq V$ is \textit{invariant} if $T_{g}\left(  W\right)
\subseteq W$ for all $g\in G$. And again, it is the minimal invariant
subspaces, the \textit{irreducible subspaces}, that are of interest. The
irreducible subspaces $\left\{  W_{\alpha}\right\}  $ are the carriers for the
\textit{irreducible representations} $T\upharpoonright:W_{\alpha}\rightarrow
W_{\alpha}$ or \textit{irreps}. And the representation space $V$ is a direct
sum of some set of irreducible subspaces $V=\sum_{i=1}^{l}\oplus W_{i}$ so the
vector-space representation of a group defines a vector-space partition of the
space. But these vector-space partitions are not unique and are thus not
canonically defined by the representation.

\begin{quotation}
\noindent Finding such a decomposition [of irreps] is an exact analogue of
finding a basis of eigenvectors of a single operator. In neither case is the
decomposition unique. However, in the operator case the eigenvalues and the
multiplicities of occurrence are uniquely determined. Moreover the linear span
of these basis vectors with a common eigenvalue is just the total eigenspace
for that eigenvalue and is uniquely determined. The decomposition as a direct
sum of eigenspaces is unique. \cite[p. 244]{mackey:wigner}
\end{quotation}

\noindent Hence the problem in this bottom-up approach is "finding an analogue
for equality of eigenvalues" \cite[p. 244]{mackey:wigner} to group the irreps together.

Suppose $T$ is a representation of $G$ acting on a space $V$ and $T^{\prime}$
is a representation of the same $G$ acting on a space $V^{\prime}$. Then a
linear map $\phi:V\rightarrow V^{\prime}$ is said \textit{morphism of
representations} or \textit{intertwining map} if for all $g\in G$ and all
$v\in V$:

\begin{center}
$\phi\left(  T_{g}\left(  v\right)  \right)  =T_{g}^{\prime}\left(
\phi\left(  v\right)  \right)  $, i.e.,

$%
\begin{array}
[c]{rcl}%
V & \overset{T_{g}}{\longrightarrow} & V\\
\phi\downarrow &  & \downarrow\phi\\
V^{\prime} & \overset{T_{g}^{\prime}}{\longrightarrow} & V^{\prime}%
\end{array}
$

commutes.
\end{center}

\noindent If $\phi$ is also invertible, then $\phi$ is said to be an
\textit{isomorphism} of representations, and the representations are said to
be \textit{isomorphic} or \textit{equivalent}.

The remarkable fact is that each group has a fixed set of inequivalent irreps,
so the distinct irreps are a characteristic of the group itself, not of a
particular representation.

The uniqueness and canonical nature of the partition obtained in the operator
case by equality of eigenvalues is now obtained using equivalence of irreps
and their underlying irreducible subspaces. All the irreducible subspaces
$W_{i}$ for irreps equivalent to an irrep $L$ in any such direct sum
$V=\sum_{i=1}^{l}\oplus W_{i}$ are grouped together (by direct sum) to obtain
the invariant carrier $W_{L}$ for a primary representation--where a
representation is \textit{primary} if all its irreducible subrepresentations
are equivalent and the underlying carrier space is also called
\textit{primary}. Note that some of the inequivalent irreps of the group $G$
may not be involved in a particular representation. The decomposition of $V$
as the direct sum $\sum_{L}\oplus W_{L}$ of the invariant primary subspaces
for the primary representations is unique. "It is the invariant subspaces
[$W_{L}$] which are the analogues of the eigenspaces of a single operator."
\cite[p. 244]{mackey:wigner} In terms of representations rather than their
carrier subspaces, it is the unique "\textit{canonical decomposition into
primary representations}." \cite[p. 244]{mackey:wigner}

Thus we have the top-down construction of the vector space partition
$V=\sum\oplus V_{i}$ of eigenspaces $V_{i}$ given by an operator (or
vector-space attribute) and the bottom-up construction of the vector-space
partition $V=\sum\oplus W_{L}$ of the carriers $W_{L}$ for primary
representations given by a vector-space representation of a symmetry group.

The following table brings out the analogies between the top-down and
bottom-up determination of vector-space partitions.%

\begin{center}
\includegraphics[
height=2.4259in,
width=5.5981in
]%
{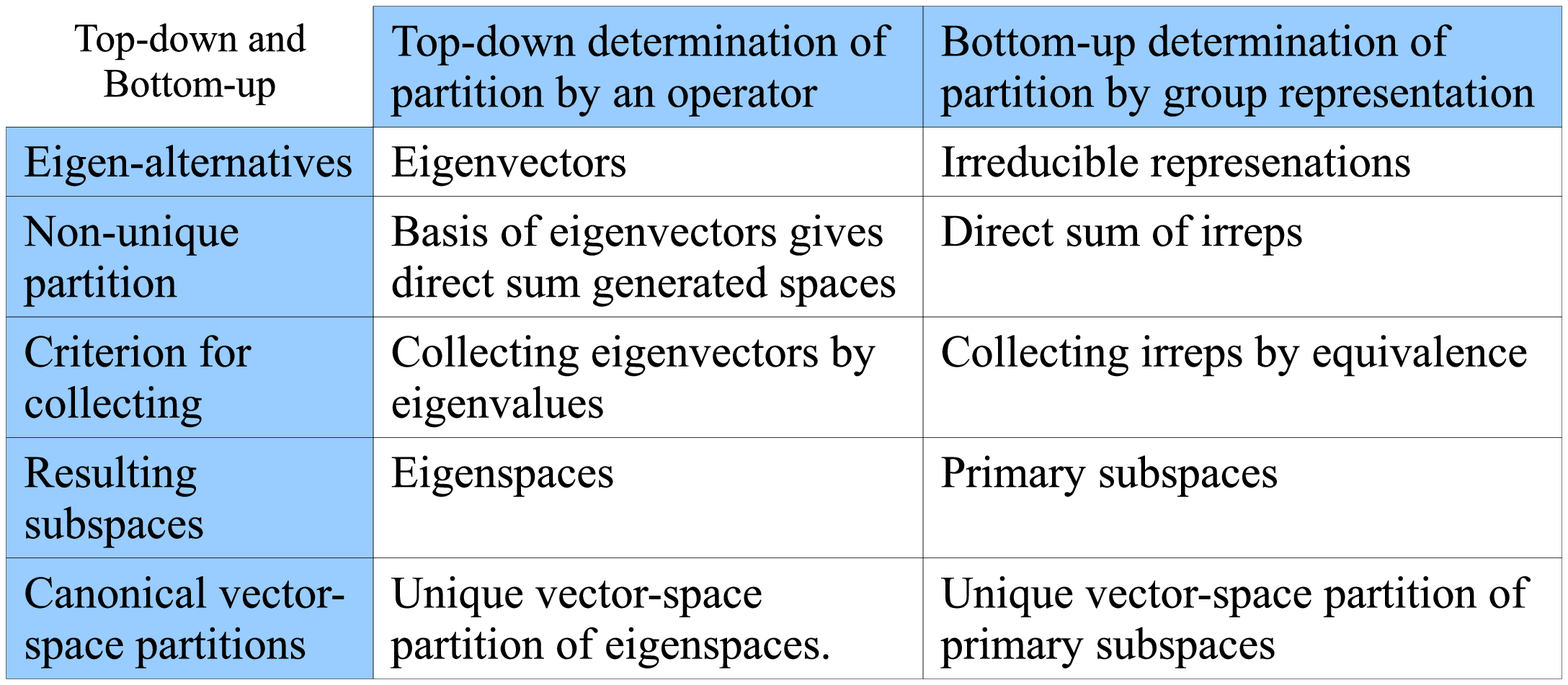}%
\\
Figure 6: Analogies between top-down and bottom-up determination of partitions
\end{center}

To represent indefiniteness, we first need to specify the "universe" of fully
definite eigen-alternatives, and then indefiniteness can be described by
collecting together or superposing the "potential" eigen-alternatives. In the
vector-space case, the eigen-alternatives determined by an operator are the
eigenvectors and the eigen-alternatives determined by a representation of a
symmetry group are the minimal invariant subspaces that are the carriers for
the irreducible representations of the symmetry group.

For state-dependent (or extrinsic) attributes of a quantum particle like the
linear momentum or angular momentum, the fully definite eigenstates are
determined by the irreducible representations of the linear-translation or
rotational-translation symmetry groups respectively. For the state-independent
(or intrinsic) attributes of quantum particles, like the mass, charge, and
spin of an electron, they are determined in particle physics by the
irreducible representations of the appropriate symmetry groups.\footnote{The
classic paper in this group-theoretic treatment of particles is Wigner
\cite{wig:unitaryirreps}. For recent overviews, see the group-theoretical
definition of particles in Falkenburg \cite{falkenburg:particlemeta} or
Roberts \cite{roberts:gsr}.}

\subsection{Vector-space partitions from other vector-space partitions}

The set notion of compatibility lifts to vector spaces, via the basis
principle, by defining two vector space partitions $\omega=\left\{
W_{\lambda}\right\}  $ and $\xi=\{X_{\mu}\}$ on $V$ as being
\textit{compatible} if there is a basis set for $V$ so that the two vector
space partitions are generated by two set partitions of that common or
simultaneous basis set.

If two vector space partitions $\omega=\left\{  W_{\lambda}\right\}  $ and
$\xi=\{X_{\mu}\}$ are compatible, then their \textit{vector space join}
$\omega\vee\xi$ is defined as the vector space partition whose subspaces are
the non-zero intersections $W_{\lambda}\cap X_{\mu}$. And by the definition of
compatibility, we could also generate the subspaces of the join $\omega\vee
\xi$ by the blocks in the set join of the two set partitions of the common
basis set.

Since real-valued set attributes lift to Hermitian linear operators, the
notion of compatible set attributes just defined would lift to two linear
operators being \textit{compatible }if their eigenspace partitions are
compatible. It is a standard fact of linear algebra \cite[p. 177]{hk:la} that
two diagonalizable linear operators $L,M:V\rightarrow V$ (on a finite
dimensional space $V$) are compatible (i.e., have a basis of simultaneous
eigenvectors) if and only if they commute, $LM=ML$. Hence the
\textit{commutativity} of linear operators is the lift of the compatibility
(i.e., defined on the same set) of set attributes. That explains the
importance of the notion of commutativity in QM and that is why the repeated
compatible measurements, described mathematically as the join operation,
requires commutativity. The join of two operator-eigenspace partitions is
defined iff the operators commute. As Weyl put it: "Thus combination [DE:
join] of two gratings [eigenspace partitions of two operators] presupposes
commutability...". \cite[p. 257]{weyl:phil}

Two commuting Hermitian linear operators $L$ and $M$ have compatible
eigenspace partitions $W_{L}=\left\{  W_{\lambda}\right\}  $ (for the
eigenvalues $\lambda$ of $L$) and $W_{M}=\left\{  W_{\mu}\right\}  $ (for the
eigenvalues $\mu$ of $M$). The blocks in the join $W_{L}\vee W_{M}$ of the two
compatible eigenspace partitions are the non-zero subspaces $\left\{
W_{\lambda}\cap W_{\mu}\right\}  $ which can be characterized by the ordered
pairs of eigenvalues $\left(  \lambda,\mu\right)  $. The nonzero vectors $v\in
W_{\lambda}\cap W_{\mu}$ are \textit{simultaneous eigenvectors} for the two
commuting operators, and there is a basis for the space consisting of
simultaneous eigenvectors.\footnote{One must be careful not to assume that the
simultaneous eigenvectors are the eigenvectors for the operator $LM=ML$ due to
the problem of degeneracy.}

A set of commuting linear operators is said to be \textit{complete} if the
join of their eigenspace partitions is nondegenerate, i.e., the blocks have
dimension $1$. The join operation gives the results of compatible measurements
so the join of a complete set of compatible vector space attributes (i.e.,
commuting Hermitian operators) gives the possible results of a non-degenerate
measurement. The eigenvectors that generate those one-dimensional blocks of
the join are characterized by the ordered $n$-tuples $\left(  \lambda
,...,\mu\right)  $ of eigenvalues so the eigenvectors are usually denoted as
the eigenkets $\left\vert \lambda,...,\mu\right\rangle $ in the Dirac
notation. These \textit{Complete Sets of Commuting Operators} are Dirac's
CSCOs \cite{dirac:principles} (which are the vector space version of our
previous CSCAs).

Since the eigen-alternatives determined by an operator, i.e., eigenvectors,
can be obtained by the complete partition joins defined by a CSCO, one might
ask if the eigen-alternatives determined by a group representation, i.e., the
irreps and their irreducible carrier spaces, could also be obtained by the
partition joins defined by some CSCO. Jin-Quan Chen and his colleagues in the
Nanjing School have developed a little-known CSCO method to systematically
find the irreducible basis vectors for the irreducible spaces that works not
only for all representations of finite groups but for all compact Lie groups
as needed in QM (\cite{chen:1985}, \cite{chen:text}). "[T]he foundation of the
new approach is precisely the theory of the complete set of commuting
operators (CSCO) initiated by Dirac..." \cite[p. 2]{chen:text} Thus the
linearized partition joins of the CSCO method extends also to all compact
group representations to characterize the maximally definite eigen-alternatives.

The partitional mathematics for sets and vector spaces is summarized in the
following table.%

\begin{center}
\includegraphics[
height=3.4877in,
width=4.6932in
]%
{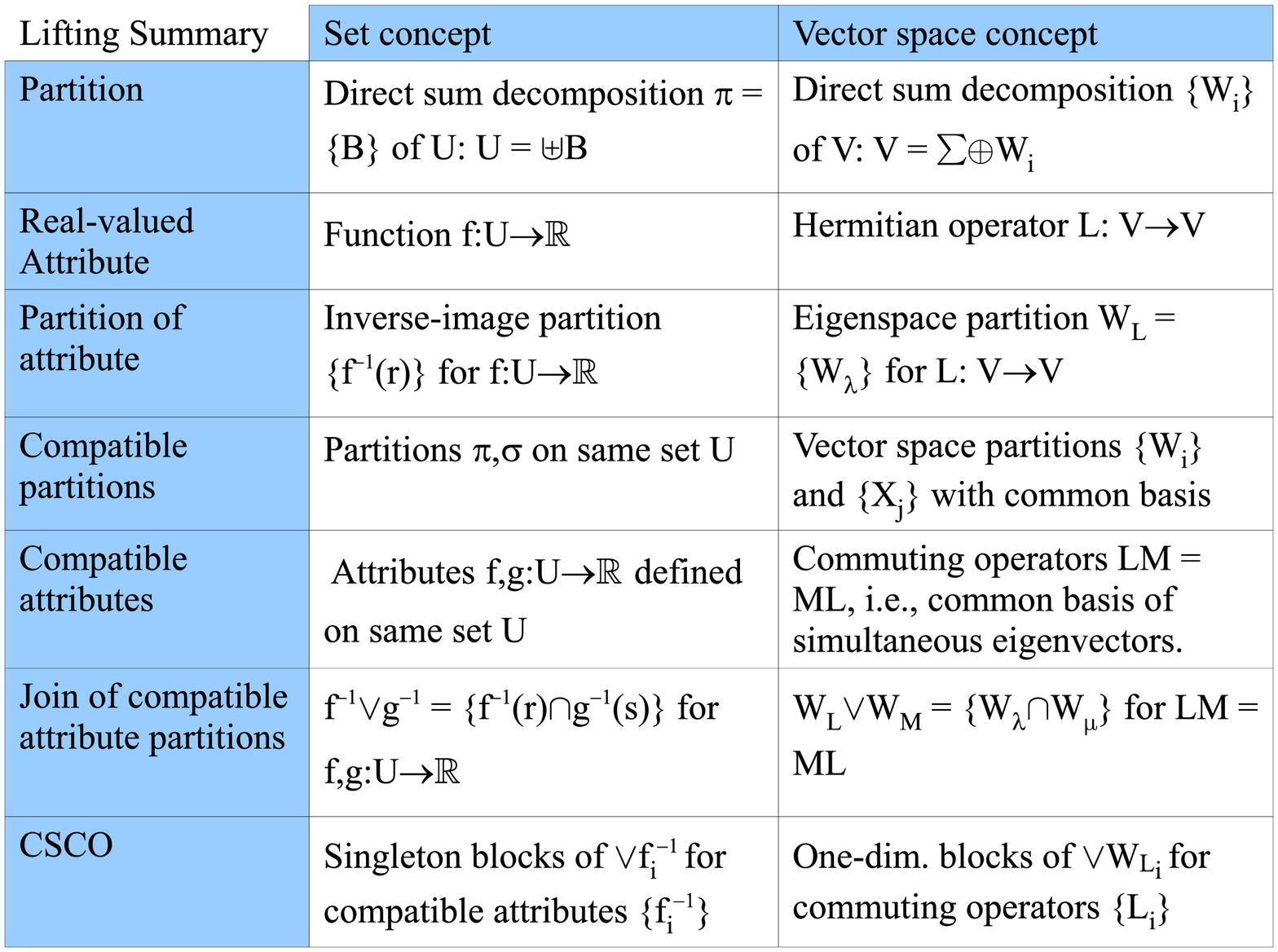}%
\\
Figure 7: Summary of partition concepts for sets and vector spaces
\end{center}

\section{Quantum Mechanics over sets}

\subsection{The pedagogical model of QM over $%
\mathbb{Z}
_{2}$}

In the tradition of toy models for quantum mechanics, Schumacher and
Westmoreland \cite{schum:modal} have recently investigated models of quantum
mechanics over finite fields. One finite field stands out over the rest, $%
\mathbb{Z}
_{2}$, since vectors in a vector space over $%
\mathbb{Z}
_{2}$ have a natural interpretation, namely as \textit{sets} that are subsets
of a universe set. But in any vector space over a finite field, there is no
inner product so the first problem in constructing a model of QM in this
context is the definition of Dirac's brackets. Which aspects of the usual
treatment of the brackets should be retained and which aspects should be dropped?

Schumacher and Westmoreland chose to have their brackets continue to have
values in the base field, e.g., $%
\mathbb{Z}
_{2}=\left\{  0,1\right\}  $, so their "theory does not make use of the idea
of probability"\cite[p. 919]{schum:modal} which certainly constrains the
relation to QM. Instead, the values of $0$ and $1$ are respectively
interpreted modally as\textit{\ impossible} and \textit{possible} and hence
their name of "modal quantum theory." A number of results from full QM carry
over to their modal quantum theory, e.g., no-cloning, superdense coding, and
teleportation, but without a probability calculus, the connection to full QM
is rather limited. And important results such as Bell's Theorem do not carry
over; "in the absence of probabilities and expectation values the Bell
approach will not work." \cite[p. 921]{schum:modal}

But all these limitations can be overcome by the different treatment of the
brackets based on crossing the sets-to-vector-space bridge in the other
direction (essentially using the basis principle in reverse). That yields a
full probability calculus for a model of \textit{quantum mechanics over sets}
(QM/sets) using the $%
\mathbb{Z}
_{2}$ base field. QM/sets yields a probability calculus--and it is a familiar
calculus, logical probability theory for a finite universe set of outcomes
developed by Laplace, Boole, and others. The only difference from that
classical calculus is the vector space formulation which allows different
(equicardinal) bases or universe sets of outcomes and thus it is the
non-commutative version of classical logical finite probability theory. This
allows the development of the QM/sets version of QM results such as Bell's
Theorem, the indeterminacy principle, double-slit experiments, and much else
in the clear and distinct context of finite sets.\footnote{Since the
development of "categorical quantum mechanics"
(\cite{abram-coecke:catprotocals} and \cite{sel:daggercompact}), it is known
that much of the mathematics of QM can be formally developed in $FdVec_{K}$,
the category of finite-dimensional vector spaces of a field $K$ and linear
maps. It is thus tempting to expect that QM/sets will be the special case of
$K=%
\mathbb{Z}
_{2}$. But this is not the case for a variety of reasons; the brackets in
QM/sets take values not in $%
\mathbb{Z}
_{2}$ but in the non-negative integers, and the attributes take their
"eigenvalues" in the reals, e.g., real-valued random variables. That is how
QM/sets turns out to be the non-commutative version of Laplace-Boole logical
finite probability theory. That is very different from a formal model of QM
where the scalars (e.g., values of brackets and eigenvalues) are in $%
\mathbb{Z}
_{2}$--such as Schumacher and Westmoreland's modal QT over $%
\mathbb{Z}
_{2}$ \cite{schum:modal}.}

By developing a sets-version of QM, the concepts and relationships of full QM
are represented in a pared-down ultra-simple version that can be seen as
representing the essential "logic" of QM. It represents the "logic of QM" in
that old sense of "logic" as giving the basic essentials of a theory, not in
the sense of giving the behavior of propositions in a theory which is the
usual "quantum logic" \cite{birkvn:logicofqm} that was, in effect, based on
the usual misdescription of Boolean subset logic as the special case of
"propositional" logic.

\subsection{Vector spaces over $%
\mathbb{Z}
_{2}$}

QM/sets is said to be "over $%
\mathbb{Z}
_{2}$" or "over sets" since the power set $\wp\left(  U\right)  \cong%
\mathbb{Z}
_{2}^{\left\vert U\right\vert }$ (for a finite non-empty universe set $U$) is
a vector space over $%
\mathbb{Z}
_{2}=\left\{  0,1\right\}  $ where the subset addition $S+T$ is the
\textit{symmetric difference} (or inequivalence) of subsets, i.e.,
$S+T=S\not \equiv T=S\cup T-S\cap T$ for $S,T\subseteq U$. Given a finite
universe set $U=\left\{  u_{1},...,u_{n}\right\}  $ of cardinality $n$, the
$U$-basis in $\wp\left(  U\right)  $ is the set of singletons $\left\{
u_{1}\right\}  ,\left\{  u_{2}\right\}  ,...,\left\{  u_{n}\right\}  $ and a
vector in $\wp\left(  U\right)  $ is specified in the $U$-basis by its $%
\mathbb{Z}
_{2}$-valued characteristic function $\chi_{S}:U\rightarrow%
\mathbb{Z}
_{2}$ for an subset $S\subseteq U$ (e.g., a string of $n$ binary numbers).
Similarly, a vector $v$ in $%
\mathbb{C}
^{n}$ is specified in terms of an orthonormal basis $\left\{  \left\vert
v_{i}\right\rangle \right\}  $ by a $%
\mathbb{C}
$-valued function $\left\langle \_|v\right\rangle :\left\{  v_{i}\right\}
\rightarrow%
\mathbb{C}
$ assigning a complex amplitude $\left\langle v_{i}|v\right\rangle $ to each
basis vector. One of the key pieces of mathematical machinery in QM, namely
the inner product, does not exist in vector spaces over finite fields but
brackets can still be defined using $\left\langle \left\{  u_{i}\right\}
|_{U}S\right\rangle =\chi_{S}\left(  u_{i}\right)  $ (see below) and a norm
can be defined to play a similar role in the probability algorithm of QM/sets.

Seeing $\wp\left(  U\right)  $ as the abstract vector space $%
\mathbb{Z}
_{2}^{n}$ allows different bases in which the vectors can be expressed (as
well as the basis-free notion of a vector as a ket). Consider the simple case
of $U=\left\{  a,b,c\right\}  $ where the $U$-basis is $\left\{  a\right\}  $,
$\left\{  b\right\}  $, and $\left\{  c\right\}  $. But the three subsets
$\left\{  a,b\right\}  $, $\left\{  b,c\right\}  $, and $\left\{
a,b,c\right\}  $ also form a basis since: $\left\{  a,b\right\}  +\left\{
a,b,c\right\}  =\left\{  c\right\}  $; $\left\{  b,c\right\}  +\left\{
c\right\}  =\left\{  b\right\}  $; and $\left\{  a,b\right\}  +\left\{
b\right\}  =\left\{  a\right\}  $. These new basis vectors could be considered
as the basis-singletons in another equicardinal universe $U^{\prime}=\left\{
a^{\prime},b^{\prime},c^{\prime}\right\}  $ where $a^{\prime}=\left\{
a,b\right\}  $, $b^{\prime}=\left\{  b,c\right\}  $, and $c^{\prime}=\left\{
a,b,c\right\}  $.

In the following \textit{ket table}, each row is a ket of $%
\mathbb{Z}
_{2}^{\left\vert U\right\vert }\cong%
\mathbb{Z}
_{2}^{3}$ expressed in the $U$-basis, the $U^{\prime}$-basis, and a
$U^{\prime\prime}$-basis.

\begin{center}%
\begin{tabular}
[c]{|c|c|c|}\hline
$U=\left\{  a,b,c\right\}  $ & $U^{\prime}=\left\{  a^{\prime},b^{\prime
},c^{\prime}\right\}  $ & $U^{\prime\prime}=\left\{  a^{\prime\prime
},b^{\prime\prime},c^{\prime\prime}\right\}  $\\\hline\hline
$\left\{  a,b,c\right\}  $ & $\left\{  c^{\prime}\right\}  $ & $\left\{
a^{\prime\prime},b^{\prime\prime},c^{\prime\prime}\right\}  $\\\hline
$\left\{  a,b\right\}  $ & $\left\{  a^{\prime}\right\}  $ & $\left\{
b^{\prime\prime}\right\}  $\\\hline
$\left\{  b,c\right\}  $ & $\left\{  b^{\prime}\right\}  $ & $\left\{
b^{\prime\prime},c^{\prime\prime}\right\}  $\\\hline
$\left\{  a,c\right\}  $ & $\left\{  a^{\prime},b^{\prime}\right\}  $ &
$\left\{  c^{\prime\prime}\right\}  $\\\hline
$\left\{  a\right\}  $ & $\left\{  b^{\prime},c^{\prime}\right\}  $ &
$\left\{  a^{\prime\prime}\right\}  $\\\hline
$\left\{  b\right\}  $ & $\left\{  a^{\prime},b^{\prime},c^{\prime}\right\}  $
& $\left\{  a^{\prime\prime},b^{\prime\prime}\right\}  $\\\hline
$\left\{  c\right\}  $ & $\left\{  a^{\prime},c^{\prime}\right\}  $ &
$\left\{  a^{\prime\prime},c^{\prime\prime}\right\}  $\\\hline
$\emptyset$ & $\emptyset$ & $\emptyset$\\\hline
\end{tabular}

Vector space isomorphism: $%
\mathbb{Z}
_{2}^{3}\cong\wp\left(  U\right)  \cong\wp\left(  U^{\prime}\right)  \cong%
\wp\left(  U^{\prime\prime}\right)  $ where row = ket.
\end{center}

\subsection{The brackets}

In a Hilbert space, the inner product is used to define the amplitudes
$\left\langle v_{i}|v\right\rangle $ and the norm $\left\vert v\right\vert
=\sqrt{\left\langle v|v\right\rangle }$ where the probability algorithm can be
formulated using this norm. In a vector space over $%
\mathbb{Z}
_{2}$, the Dirac notation can still be used to define a norm even though there
is no inner product. For a singleton basis element $\left\{  u\right\}
\subseteq U$, the \textit{bra} $\left\langle \left\{  u\right\}  \right\vert
_{U}:\wp\left(  U\right)  \rightarrow%
\mathbb{R}
$ is defined by the \textit{bracket}:

\begin{center}
$\left\langle \left\{  u\right\}  |_{U}S\right\rangle =\left\{
\begin{array}
[c]{c}%
1\text{ if }u\in S\\
0\text{ if }u\notin S
\end{array}
\right.  =\chi_{S}\left(  u\right)  $
\end{center}

\noindent Then $\left\langle \left\{  u_{i}\right\}  |_{U}\left\{
u_{j}\right\}  \right\rangle =\chi_{\left\{  u_{j}\right\}  }\left(
u_{i}\right)  =\chi_{\left\{  u_{i}\right\}  }\left(  u_{j}\right)
=\delta_{ij}$ is the set-version of $\left\langle v_{i}|v_{j}\right\rangle
=\delta_{ij}$ (for an orthonormal basis $\left\{  \left\vert v_{i}%
\right\rangle \right\}  $). Assuming a finite $U$, the bracket linearly
extends to the more general form (where $\left\vert S\right\vert $ is the
cardinality of $S$):

\begin{center}
$\left\langle T|_{U}S\right\rangle =\left\vert T\cap S\right\vert $ for
$T,S\subseteq U$.\footnote{Thus $\left\langle T|_{U}S\right\rangle =\left\vert
T\cap S\right\vert $ takes values outside the base field of $%
\mathbb{Z}
_{2}$ just like the Hamming distance function $d_{H}\left(  T,S\right)
=\left\vert T+S\right\vert $ on vector spaces over $%
\mathbb{Z}
_{2}$ in coding theory. \cite{mceliece:coding}}
\end{center}

The basis principle can be run in reverse to transport a vector space concept
to sets. Consider an orthonormal basis set $\left\{  \left\vert v_{i}%
\right\rangle \right\}  $ in a finite dimensional Hilbert space. Given two
subsets $T,S\subseteq\left\{  \left\vert v_{i}\right\rangle \right\}  $ of the
basis set, consider the unnormalized superpositions $\psi_{T}=\sum_{\left\vert
v_{i}\right\rangle \in T}\left\vert v_{i}\right\rangle $ and $\psi_{S}%
=\sum_{\left\vert v_{i}\right\rangle \in S}\left\vert v_{i}\right\rangle $.
Then their inner product in the Hilbert space is $\left\langle \psi_{T}%
|\psi_{S}\right\rangle =\left\vert T\cap S\right\vert $, which transports
(crossing the bridge in the other direction) to $\left\langle T|_{U}%
S\right\rangle =\left\vert T\cap S\right\vert $ for subsets $T,S\subseteq U$
of the $U$-basis of $%
\mathbb{Z}
_{2}^{\left\vert U\right\vert }$. In both cases, the bracket gives the size of
the overlap or indistinctness of the two vectors or sets.

\subsection{Ket-bra resolution}

The \textit{ket-bra} $\left\vert \left\{  u\right\}  \right\rangle
\left\langle \left\{  u\right\}  \right\vert _{U}$ is the one-dimensional
projection operator:

\begin{center}
$\left\vert \left\{  u\right\}  \right\rangle \left\langle \left\{  u\right\}
\right\vert _{U}=\left\{  u\right\}  \cap():\wp\left(  U\right)
\rightarrow\wp\left(  U\right)  $
\end{center}

\noindent and the \textit{ket-bra identity} holds as usual:

\begin{center}
$\sum_{u\in U}\left\vert \left\{  u\right\}  \right\rangle \left\langle
\left\{  u\right\}  \right\vert _{U}=\sum_{u\in U}\left(  \left\{  u\right\}
\cap()\right)  =I:\wp\left(  U\right)  \rightarrow\wp\left(  U\right)  $
\end{center}

\noindent where the summation is the symmetric difference of sets in
$\wp\left(  U\right)  $. The overlap $\left\langle T|_{U}S\right\rangle $ can
be resolved using the ket-bra identity in the same basis: $\left\langle
T|_{U}S\right\rangle =\sum_{u}\left\langle T|_{U}\left\{  u\right\}
\right\rangle \left\langle \left\{  u\right\}  |_{U}S\right\rangle $.
Similarly a ket $\left\vert S\right\rangle $ for $S\subseteq U$ can be
resolved in the $U$-basis;

\begin{center}
$\left\vert S\right\rangle =\sum_{u\in U}\left\vert \left\{  u\right\}
\right\rangle \left\langle \left\{  u\right\}  |_{U}S\right\rangle =\sum_{u\in
U}\left\langle \left\{  u\right\}  |_{U}S\right\rangle \left\vert \left\{
u\right\}  \right\rangle =\sum_{u\in U}\left\vert \left\{  u\right\}  \cap
S\right\vert \left\vert \left\{  u\right\}  \right\rangle $
\end{center}

\noindent where a subset $S\subseteq U$ is just expressed as the sum of the
singletons $\left\{  u\right\}  \subseteq S$. That is ket-bra resolution in
sets. The ket $\left\vert S\right\rangle $ is the same as the ket $\left\vert
S^{\prime}\right\rangle $ for some subset $S^{\prime}\subseteq U^{\prime}$ in
another $U^{\prime}$-basis, but when the bra $\left\langle \left\{  u\right\}
\right\vert _{U}$ is applied to the ket $\left\vert S\right\rangle =\left\vert
S^{\prime}\right\rangle $, then it is the subset $S\subseteq U$, not
$S^{\prime}\subseteq U^{\prime}$, that comes outside the ket symbol
$\left\vert \ \right\rangle $ in $\left\langle \left\{  u\right\}
|_{U}S\right\rangle =\left\vert \left\{  u\right\}  \cap S\right\vert
$.\footnote{The term "$\left\{  u\right\}  \cap S^{\prime}$" is not even
defined since it is the intersection of subsets of two different universes. }

\subsection{The norm}

The $U$\textit{-norm} $\left\Vert S\right\Vert _{U}:\wp\left(  U\right)
\rightarrow%
\mathbb{R}
$ is defined, as usual, as the square root of the bracket:\footnote{We use the
double-line notation $\left\Vert S\right\Vert _{U}$ for the $U$-norm of a set
to distinguish it from the single-line notation $\left\vert S\right\vert $ for
the cardinality of a set, whereas the customary absolute value notation for
the norm of a vector in full QM is $\left\vert v\right\vert $.}

\begin{center}
$\left\Vert S\right\Vert _{U}=\sqrt{\left\langle S|_{U}S\right\rangle }%
=\sqrt{|S|}$
\end{center}

\noindent for $S\in\wp\left(  U\right)  $ which is the set-version of the norm
$\left\vert \psi\right\vert =\sqrt{\left\langle \psi|\psi\right\rangle }$.
Note that a ket has to be expressed in the $U$-basis to apply the norm
definition so in the above example, $\left\Vert \left\{  a^{\prime}\right\}
\right\Vert _{U}=\sqrt{2}$ since $\left\{  a^{\prime}\right\}  =\left\{
a,b\right\}  $ in the $U$-basis.

\subsection{The Born Rule}

For a specific basis $\left\{  \left\vert v_{i}\right\rangle \right\}  $ and
for any nonzero vector $v$ in a finite dimensional complex vector space,
$\left\vert v\right\vert ^{2}=\sum_{i}\left\langle v_{i}|v\right\rangle
\left\langle v_{i}|v\right\rangle ^{\ast}$ ($^{\ast}$ is complex conjugation)
whose set version would be: $\left\Vert S\right\Vert _{U}^{2}=\sum_{u\in
U}\left\langle \left\{  u\right\}  |_{U}S\right\rangle ^{2} $. Since

\begin{center}
$\left\vert v\right\rangle =\sum_{i}\left\langle v_{i}|v\right\rangle
\left\vert v_{i}\right\rangle $ and $\left\vert S\right\rangle =\sum_{u\in
U}\left\langle \left\{  u\right\}  |_{U}S\right\rangle \left\vert \left\{
u\right\}  \right\rangle $,
\end{center}

\noindent applying the Born Rule by squaring the coefficients $\left\langle
v_{i}|v\right\rangle $ and $\left\langle \left\{  u\right\}  |_{U}%
S\right\rangle $ (and normalizing) gives the probability sums for the
eigen-elements $v_{i}$ or $\left\{  u\right\}  $ given a state $v$ or $S$
respectively in QM and QM/sets:

\begin{center}
$\sum_{i}\frac{\left\langle v_{i}|v\right\rangle \left\langle v_{i}%
|v\right\rangle ^{\ast}}{\left\vert v\right\vert ^{2}}=1$ and $\sum_{u}%
\frac{\left\langle \left\{  u\right\}  |_{U}S\right\rangle ^{2}}{\left\Vert
S\right\Vert _{U}^{2}}=\sum_{u}\frac{\left\vert \left\{  u\right\}  \cap
S\right\vert }{\left\vert S\right\vert }=1$
\end{center}

\noindent where $\frac{\left\langle v_{i}|v\right\rangle \left\langle
v_{i}|v\right\rangle ^{\ast}}{\left\vert v\right\vert ^{2}}$ is a `mysterious'
quantum probability while $\frac{\left\langle \left\{  u\right\}
|_{U}S\right\rangle ^{2}}{\left\Vert S\right\Vert _{U}^{2}}=\frac{\left\vert
\left\{  u\right\}  \cap S\right\vert }{\left\vert S\right\vert }$ is the
unmysterious Laplacian equal probability $\Pr\left(  \left\{  u\right\}
|S\right)  $ rule for getting $u$ when sampling $S$.\footnote{Note that there
is no notion of a normalized vector in a vector space over $%
\mathbb{Z}
_{2}$ (another consequence of the lack of an inner product). The normalization
is, as it were, postponed to the probability algorithm which is computed in
the rationals.}

\subsection{Spectral decomposition on sets}

An observable, i.e., a Hermitian operator, on a Hilbert space has a home basis
set of orthonormal eigenvectors. In a similar manner, a real-valued attribute
$f:U\rightarrow%
\mathbb{R}
$ defined on $U$ has the $U$-basis as its "home basis set." The connection
between the numerical attributes $f:U\rightarrow%
\mathbb{R}
$ of QM/sets and the Hermitian operators of full QM can be established by
"seeing" the function $f$ as a formal operator: $f\upharpoonright():\wp\left(
U\right)  \rightarrow\wp\left(  U\right)  $. Applied to the basis elements
$\left\{  u\right\}  \subseteq U$, we may write $f\upharpoonright\left\{
u\right\}  =f\left(  u\right)  \left\{  u\right\}  =r\left\{  u\right\}  $ as
the set-version of an eigenvalue equation applied to an eigenvector, where the
multiplication $r\left\{  u\right\}  $ is only formal (read $r\left\{
u\right\}  $ as the instruction: \texttt{give }$f$\texttt{ the value }%
$r$\texttt{ on }$\left\{  u\right\}  $). Then for any subset $S\subseteq
f^{-1}\left(  r\right)  $ where $f$ is constant, we may also formally write:
$f\upharpoonright S=rS$ as an eigenvalue equation satisfied by all the
eigen-sets or eigenvectors $S$ in the eigenspace $\wp\left(  f^{-1}\left(
r\right)  \right)  $, a subspace of $\wp\left(  U\right)  $, for the
eigenvalue $r$. The eigenspaces$\wp\left(  f^{-1}\left(  r\right)  \right)  $
give a direct sum decomposition (i.e., a vector-space partition) of the whole
space $\wp\left(  U\right)  =\sum_{r}\oplus\wp\left(  f^{-1}\left(  r\right)
\right)  $, just as the set partition $f^{-1}=\left\{  f^{-1}\left(  r\right)
\right\}  _{r}$ gives a direct sum decomposition of the set $U=%
{\textstyle\biguplus\nolimits_{r}}
f^{-1}\left(  r\right)  $. Since $f^{-1}\left(  r\right)  \cap():\wp\left(
U\right)  \rightarrow\wp\left(  U\right)  $ is the projection
operator\footnote{Since $\wp\left(  U\right)  $ is now interpreted as a vector
space, it should be noted that the projection operator $T\cap():\wp\left(
U\right)  \rightarrow\wp\left(  U\right)  $ is not only idempotent but linear,
i.e., $\left(  T\cap S_{1}\right)  +(T\cap S_{2})=T\cap\left(  S_{1}%
+S_{2}\right)  $. Indeed, this is the distributive law when $\wp\left(
U\right)  $ is interpreted as a Boolean ring.} to the eigenspace$\wp\left(
f^{-1}\left(  r\right)  \right)  $ for the eigenvalue $r$, we have the
spectral decomposition of a $U$-attribute $f:U\rightarrow%
\mathbb{R}
$ in QM/sets analogous to the spectral decomposition of a Hermitian operator
$L=\sum_{\lambda}\lambda P_{\lambda}$ in QM:

\begin{center}
$f\upharpoonright()=\sum_{r}r\left[  f^{-1}\left(  r\right)  \cap()\right]
:\wp\left(  U\right)  \rightarrow\wp\left(  U\right)  $

$L=\sum_{\lambda}\lambda P_{\lambda}:V\rightarrow V$

Spectral decomposition of operators in QM/sets and in QM.
\end{center}

\subsection{Lifting and internalization}

QM/sets is a pedagogical model where the internal workings of QM math are laid
out in a simplified and "externalized" form. Think of a simplified laboratory
model of a complex machine where the parts are visible and easily laid out to
clarify the workings of the actual machine. To recover QM math from QM/sets,
the base field is lifted from the sets-case of $%
\mathbb{Z}
_{2}$ to QM-case of $%
\mathbb{C}
$, and the externalized forms become \textit{internalized} (or "encoded")
within the vector space. The formal multiplication $r\left[  f^{-1}\left(
r\right)  \cap()\right]  $ is internalized as an actual multiplication of a
scalar times an operator $\lambda P_{\lambda}$ on a vector space over $%
\mathbb{C}
$. The operator representation $L=\sum_{\lambda}\lambda P_{\lambda}$ of an
observable is just the lifted internalization or encoding of a numerical
attribute $\sum_{r}r\left[  f^{-1}\left(  r\right)  \cap()\right]  $ made
possible by the enriched base field $%
\mathbb{C}
$. The set brackets $\left\langle S|_{U}T\right\rangle $ taking values outside
the base field $%
\mathbb{Z}
_{2}$ become internalized as an inner product with the same enrichment of the
base field to $%
\mathbb{C}
$, and similarly for the $U$-norm that is the square root of the
brackets.\footnote{The Schumacher-Westmoreland decision to try to develop
quantum theory over $%
\mathbb{Z}
_{2}$ with the brackets taking values in the base field $%
\mathbb{Z}
_{2}$ \cite{schum:modal} can thus be seen as an example of premature
internalization.}

It is the comparative poverty of the base field $%
\mathbb{Z}
_{2}$ that requires the QM/sets brackets and norm to take the externalized
values outside the base field and for a formal multiplication to be used in
the "operator" representation $f\upharpoonright()=\sum_{r}r\left[
f^{-1}\left(  r\right)  \cap()\right]  $ of a numerical attribute
$f:U\rightarrow%
\mathbb{R}
$. The only numerical attributes that can be internally represented in
$\wp\left(  U\right)  \cong%
\mathbb{Z}
_{2}^{\left\vert U\right\vert }$ are the $0,1$-attributes or characteristic
functions $\chi_{S}:U\rightarrow%
\mathbb{Z}
_{2}$ that are internally represented in the $U$-basis as the projection
operators $S\cap():\wp\left(  U\right)  \rightarrow\wp\left(  U\right)  $.

In the engineering literature, eigenvalues are seen as "stretching or
shrinking factors" but that is \textit{not} their role in QM. The whole
machinery of eigenvectors [e.g., the eigen-sets $f\upharpoonright S=rS$ for
$S\subseteq f^{-1}\left(  r\right)  $ in sets], eigenspaces [e.g., the space
of all eigen-sets $\wp\left(  f^{-1}\left(  r\right)  \right)  $], and
eigenvalues [e.g., $f(u)=r$] in full QM is a way of lifting and internalizing
or encoding numerical attributes [e.g., $f:U\rightarrow%
\mathbb{R}
$ in the set case] \textit{inside} a vector space that has a rich enough base
field. The old question "Why do attributes in classical physics (like the
position or momentum of a particle) become operators in QM?" is addressed by
\textit{internalization}. The observable operators of QM \textit{are} the
lifted vector space internalizations or encodings over $%
\mathbb{C}
$ of the concept of real-valued attributes (or random variables) on sets.

Moreover, for the internalization of attributes as operators to always be
possible, the secular equations for eigenvalues have to have a complete set of
solutions so the base field has to be algebraically closed--which addresses
another old question of why full QM has the complex numbers $%
\mathbb{C}
$ as its base field.

\subsection{Completeness and orthogonality of projection operators}

The usual completeness and orthogonality conditions on eigenspaces also have
set-versions in QM over $%
\mathbb{Z}
_{2}$:

\begin{enumerate}
\item completeness: $\sum_{\lambda}P_{\lambda}=I:V\rightarrow V$ has the
set-version: $\sum_{r}f^{-1}\left(  r\right)  \cap()=I:\wp\left(  U\right)
\rightarrow\wp\left(  U\right)  $, and

\item orthogonality: for $\lambda\neq\lambda^{\prime}$, $P_{\lambda}%
P_{\lambda^{\prime}}=0:V\rightarrow V$ (where $0$ is the zero operator) has
the set-version: for $r\neq r^{\prime}$, $\left[  f^{-1}\left(  r\right)
\cap()\right]  \left[  f^{-1}\left(  r^{\prime}\right)  \cap()\right]
=\emptyset\cap():\wp\left(  U\right)  \rightarrow\wp\left(  U\right)  $.
\end{enumerate}

Note that in spite of the lack of an inner product, the orthogonality of
projection operators $S\cap()$ is perfectly well defined in QM/sets where it
boils down to the disjointness of subsets, i.e., the cardinality of their
overlap (instead of their inner product) being $0$.

\subsection{Pythagorean Theorem for sets}

An orthogonal decomposition of a finite set $U$ is just a partition
$\pi=\left\{  B\right\}  $ of $U$ since the blocks $B,B^{\prime},...$ are
orthogonal (i.e., disjoint) and their sum is $U$. Given such an orthogonal
decomposition of $U$, we have the:

\begin{center}
$\left\Vert U\right\Vert _{U}^{2}=\sum_{B\in\pi}\left\Vert B\right\Vert
_{U}^{2}$

Pythagorean Theorem

for orthogonal decompositions of sets.
\end{center}

\subsection{Whence the Born Rule?}

Another old question is: "why the squaring of amplitudes in QM?" A state
objectively indefinite between certain definite orthogonal alternatives $A$
and $B$, where the latter are represented by vectors $\overrightarrow{A}$ and
$\overrightarrow{B}$, is represented by the vector sum $\overrightarrow{C}%
=\overrightarrow{A}+\overrightarrow{B}$. But what is the "strength,"
"intensity," or relative importance of the vectors $\overrightarrow{A}$ and
$\overrightarrow{B}$ in the vector sum $\overrightarrow{C}$? That question
requires a \textit{scalar} measure of strength or intensity. The magnitude
given by the norm does not answer the question since $\left\Vert
\overrightarrow{A}\right\Vert +\left\Vert \overrightarrow{B}\right\Vert
\neq\left\Vert \overrightarrow{C}\right\Vert $. But the Pythagorean Theorem
shows that the norm-squared gives the scalar measure of "intensity" that
answers the question: $\left\Vert \overrightarrow{A}\right\Vert ^{2}%
+\left\Vert \overrightarrow{B}\right\Vert ^{2}=\left\Vert \overrightarrow{C}%
\right\Vert ^{2}$ in vector spaces over $%
\mathbb{Z}
_{2}$ or over $%
\mathbb{C}
$. And when the objectively indefinite superposition state is decohered by a
distinction-making measurement, then the \textit{objective probability} that
the indefinite state will reduce to one of the definite alternatives is given
by that objective relative scalar measure of the eigen-alternative's
"strength," "intensity," or importance in the indefinite state--and that is
the Born Rule. In a slogan, Born is the son of Pythagoras.

\subsection{Measurement in QM/sets}

\noindent The Pythagorean results (for the complete and orthogonal projection operators):

\begin{center}
$\left\vert v\right\vert ^{2}=\sum_{\lambda}\left\vert P_{\lambda}\left(
v\right)  \right\vert ^{2}$ and $\left\Vert S\right\Vert _{U}^{2}=\sum
_{r}\left\Vert f^{-1}\left(  r\right)  \cap S\right\Vert _{U}^{2}$,
\end{center}

\noindent give the probabilities for measuring attributes. Since by the
Pythagorean Theorem:

\begin{center}
$\left\vert S\right\vert =\left\Vert S\right\Vert _{U}^{2}=\sum_{r}\left\Vert
f^{-1}\left(  r\right)  \cap S\right\Vert _{U}^{2}=\sum_{r}\left\vert
f^{-1}\left(  r\right)  \cap S\right\vert $,
\end{center}

\noindent we have in full QM and in QM/sets:

\begin{center}
$\sum_{\lambda}\frac{\left\vert P_{\lambda}\left(  v\right)  \right\vert ^{2}%
}{\left\vert v\right\vert ^{2}}=1$ and $\sum_{r}\frac{\left\Vert f^{-1}\left(
r\right)  \cap S\right\Vert _{U}^{2}}{\left\Vert S\right\Vert _{U}^{2}}%
=\sum_{r}\frac{\left\vert f^{-1}\left(  r\right)  \cap S\right\vert
}{\left\vert S\right\vert }=1$.
\end{center}

\noindent Here $\frac{\left\vert P_{\lambda}\left(  v\right)  \right\vert
^{2}}{\left\vert v\right\vert ^{2}}$ is the mysterious quantum probability of
getting $\lambda$ in an $L$-measurement of $v$ while $\frac{\left\vert
f^{-1}\left(  r\right)  \cap S\right\vert }{\left\vert S\right\vert }$ has the
rather unmysterious interpretation in the pedagogical model, QM/sets, as the
probability $\Pr\left(  r|S\right)  $ of the random variable $f:U\rightarrow%
\mathbb{R}
$ having the eigen-value $r$ when sampling $S\subseteq U$. Thus the
set-version of the Born Rule is not some weird quantum notion of probability
on sets but the perfectly ordinary Laplace-Boole rule for the conditional
probability $\Pr\left(  r|S\right)  =\frac{\left\vert f^{-1}\left(  r\right)
\cap S\right\vert }{\left\vert S\right\vert }$, given $S\subseteq U$, of a
random variable $f:U\rightarrow%
\mathbb{R}
$ having the value $r$.

The collecting-together of some eigen-elements $\left\{  u\right\}  \subseteq
U$ into a subset $S\subseteq U$ to form an "indefinite element" $S$ has the
vector sum $\left\vert S\right\rangle =\sum_{u\in U}\left\langle \left\{
u\right\}  |_{U}S\right\rangle \left\vert \left\{  u\right\}  \right\rangle $
in the vector space $\wp\left(  U\right)  $ over $%
\mathbb{Z}
_{2}$ giving the superposition version of the indefinite element. This cements
the interpretation of collecting together in sets as superposition in vector spaces.

The indefinite element $S$ is being measured using the observable $f$ where
the probability $\Pr\left(  r|S\right)  $ of getting the eigenvalue $r$ is
$\frac{\left\Vert f^{-1}\left(  r\right)  \cap S\right\Vert _{U}^{2}%
}{\left\Vert S\right\Vert _{U}^{2}}=\frac{\left\vert f^{-1}\left(  r\right)
\cap S\right\vert }{\left\vert S\right\vert }$ and where the "damned quantum
jump" (Schr\"{o}dinger) goes from $S$ by the projection operator
$f^{-1}\left(  r\right)  \cap()$ to the projected resultant state
$f^{-1}\left(  r\right)  \cap S$ which is in the eigenspace $\wp\left(
f^{-1}\left(  r\right)  \right)  $ for that eigenvalue $r$.

The partition operation in QM/sets that describes measurement is the partition
join of the partition $\left\{  S,S^{c}\right\}  $ and $f^{-1}=\left\{
f^{-1}\left(  r\right)  \right\}  $ so that the initial pure state $S$ (as a
mini-blob) is refined into the mixture $\left\{  f^{-1}\left(  r\right)  \cap
S\right\}  $ of possible resultant states. The other states $\left\{
f^{-1}\left(  r\right)  \cap S^{c}\right\}  $ in the join $f^{-1}\vee\left\{
S,S^{c}\right\}  $ are not possible or "potential" states starting from $S$.
The state resulting from the measurement represents a more-definite element
$f^{-1}\left(  r\right)  \cap S$ that now has the definite $f$-value of
$r$--so a second measurement would yield the same eigenvalue $r$ with
probability $\Pr\left(  r|f^{-1}\left(  r\right)  \cap S\right)
=\frac{\left\vert f^{-1}\left(  r\right)  \cap\left[  f^{-1}\left(  r\right)
\cap S\right]  \right\vert }{\left\vert f^{-1}\left(  r\right)  \cap
S\right\vert }=\frac{\left\vert f^{-1}\left(  r\right)  \cap S\right\vert
}{\left\vert f^{-1}\left(  r\right)  \cap S\right\vert }=1$ and the same
vector $f^{-1}\left(  r\right)  \cap\left[  f^{-1}\left(  r\right)  \cap
S\right]  =f^{-1}\left(  r\right)  \cap S$ using the idempotency of the
set-version of projection operators.

This is \textit{all }just the QM over sets version of the treatment of
measurement in standard Dirac-von-Neumann QM where the probability calculus of
logical probability theory is lifted and internalized by the Born Rule and
where the real attributes or real random variables $f:U\rightarrow%
\mathbb{R}
$ are internalized as observable (Hermitian) operators.\footnote{See
\cite{ell:objindef} and \cite{ell:qmoversets} for a more extensive treatment
of measurement using density matrices in both full QM and QM/sets.}

\subsection{Summary of QM/sets}

These set-versions are summarized in the following table for a finite $U$ and
a finite dimensional Hilbert space $V$ with $\left\{  \left\vert
v_{i}\right\rangle \right\}  $ as any orthonormal basis.

\begin{center}%
\begin{tabular}
[c]{|c|c|}\hline
Vector space over $%
\mathbb{Z}
_{2}$: QM/sets & Hilbert space case: QM over $%
\mathbb{C}
$\\\hline\hline
Projections: $S\cap():\wp\left(  U\right)  \rightarrow\wp\left(  U\right)  $ &
$P:V\rightarrow V$\\\hline
Spectral Decomp.: $f\upharpoonright()=\sum_{r}r\left(  f^{-1}\left(  r\right)
\cap()\right)  $ & $L=\sum_{\lambda}\lambda P_{\lambda}$\\\hline
Compl.: $\sum_{r}f^{-1}\left(  r\right)  \cap()=I:\wp\left(  U\right)
\rightarrow\wp\left(  U\right)  $ & $\sum_{\lambda}P_{\lambda}=I$\\\hline
Orthog.: $r\neq r^{\prime}$, $\left[  f^{-1}\left(  r\right)  \cap()\right]
\left[  f^{-1}\left(  r^{\prime}\right)  \cap()\right]  =\emptyset\cap()$ &
$\lambda\neq\lambda^{\prime}$, $P_{\lambda}P_{\lambda^{\prime}}=0$\\\hline
Brackets: $\left\langle S|_{U}T\right\rangle =\left\vert S\cap T\right\vert $
= overlap of $S,T\subseteq U$ & $\left\langle \psi|\varphi\right\rangle =$
overlap of $\psi$ and $\varphi$\\\hline
Ket-bra: $\sum_{u\in U}\left\vert \left\{  u\right\}  \right\rangle
\left\langle \left\{  u\right\}  \right\vert _{U}=\sum_{u\in U}\left(
\left\{  u\right\}  \cap()\right)  =I$ & $\sum_{i}\left\vert v_{i}%
\right\rangle \left\langle v_{i}\right\vert =I$\\\hline
Resolution: $\left\langle S|_{U}T\right\rangle =\sum_{u}\left\langle
S|_{U}\left\{  u\right\}  \right\rangle \left\langle \left\{  u\right\}
|_{U}T\right\rangle $ & $\left\langle \psi|\varphi\right\rangle =\sum
_{i}\left\langle \psi|v_{i}\right\rangle \left\langle v_{i}|\varphi
\right\rangle $\\\hline
Norm: $\left\Vert S\right\Vert _{U}=\sqrt{\left\langle S|_{U}S\right\rangle
}=\sqrt{\left\vert S\right\vert }$ where $S\subseteq U$ & $\left\vert
\psi\right\vert =\sqrt{\left\langle \psi|\psi\right\rangle }$\\\hline
Pythagoras: $\left\Vert S\right\Vert _{U}^{2}=\sum_{u\in U}\left\langle
\left\{  u\right\}  |_{U}S\right\rangle ^{2}=\left\vert S\right\vert $ &
$\left\vert \psi\right\vert ^{2}=\sum_{i}\left\langle v_{i}|\psi\right\rangle
^{\ast}\left\langle v_{i}|\psi\right\rangle $\\\hline
Laplace: $S\neq\emptyset$, $\sum_{u\in U}\frac{\left\langle \left\{
u\right\}  |_{U}S\right\rangle ^{2}}{\left\Vert S\right\Vert _{U}^{2}}%
=\sum_{u\in S}\frac{1}{\left\vert S\right\vert }=1$ & $\left\vert
\psi\right\rangle \neq0$, $\sum_{i}\frac{\left\langle v_{i}|\psi\right\rangle
^{\ast}\left\langle v_{i}|\psi\right\rangle }{\left\vert \psi\right\vert ^{2}%
}=\frac{\left\vert \left\langle v_{i}|\psi\right\rangle \right\vert ^{2}%
}{\left\vert \psi\right\vert ^{2}}=1$\\\hline
Born: $\left\vert S\right\rangle =\sum_{u\in U}\left\langle \left\{
u\right\}  |_{U}S\right\rangle \left\vert \left\{  u\right\}  \right\rangle $,
$\Pr\left(  \left\{  u\right\}  |S\right)  =\frac{\left\langle \left\{
u\right\}  |_{U}S\right\rangle ^{2}}{\left\Vert S\right\Vert _{U}^{2}}$ &
$\left\vert \psi\right\rangle =\sum_{i}\left\langle v_{i}|\psi\right\rangle
\left\vert v_{i}\right\rangle $, $\Pr\left(  v_{i}|\psi\right)  =\frac
{\left\vert \left\langle v_{i}|\psi\right\rangle \right\vert ^{2}}{\left\vert
\psi\right\vert ^{2}}$\\\hline
$\left\Vert S\right\Vert _{U}^{2}=\sum_{r}\left\Vert f^{-1}\left(  r\right)
\cap S\right\Vert _{U}^{2}=\sum_{r}\left\vert f^{-1}\left(  r\right)  \cap
S\right\vert =\left\vert S\right\vert $ & $\left\vert \psi\right\vert
^{2}=\sum_{\lambda}\left\vert P_{\lambda}\left(  \psi\right)  \right\vert
^{2}$\\\hline
$S\neq\emptyset$, $\sum_{r}\frac{\left\Vert f^{-1}\left(  r\right)  \cap
S\right\Vert _{U}^{2}}{\left\Vert S\right\Vert _{U}^{2}}=\sum_{r}%
\frac{\left\vert f^{-1}\left(  r\right)  \cap S\right\vert }{\left\vert
S\right\vert }=1$ & $\left\vert \psi\right\rangle \neq0$, $\sum_{\lambda}%
\frac{\left\vert P_{\lambda}\left(  \psi\right)  \right\vert ^{2}}{\left\vert
\psi\right\vert ^{2}}=1$\\\hline
Measurement: $\Pr(r|S)=\frac{\left\Vert f^{-1}\left(  r\right)  \cap
S\right\Vert _{U}^{2}}{\left\Vert S\right\Vert _{U}^{2}}=\frac{\left\vert
f^{-1}\left(  r\right)  \cap S\right\vert }{\left\vert S\right\vert }$ &
$\Pr\left(  \lambda|\psi\right)  =\frac{\left\vert P_{\lambda}\left(
\psi\right)  \right\vert ^{2}}{\left\vert \psi\right\vert ^{2}}$\\\hline
\end{tabular}

Probability mathematics for QM over $%
\mathbb{Z}
_{2}$ and for QM over $%
\mathbb{C}
$
\end{center}

Since the probability calculus on the left-side of the table is a
non-commutative version of the old Laplace-Boole \textit{logical} finite
probability theory \cite{boole:lot}, the correspondence with the right-side of
the table gives precision to the thesis that the mathematics of QM is a type
of \textit{logical probability calculus complexified in Hilbert spaces}. For
instance, the equiprobability or indifference principle of that classical
logical finite probability calculus--that is, the equality of all non-zero
coefficients in a vector over $%
\mathbb{Z}
_{2}$--generalizes to the QM probability calculus using the coefficients of
vectors over $%
\mathbb{C}
$, and the Born Rule applies to orthogonal decompositions in both cases.
Crossing the sets-vector-space bridge the other way, we have seen that the
classical logical probability calculus (in its non-commutative version) is a
type of quantum mechanics over sets.

\subsection{Whence von Neumann's Type 1 and Type 2 processes?}

When some particularly mysterious process like measurement can be clearly and
distinctly modeled in QM/sets, then it casts some sense-making light back on
full QM. A good example is von Neumann's distinction between Type 1
measurement-like processes and Type 2 processes of unitary evolution
\cite{vonn:mfqm}. In QM/sets, we have seen that measurement is a
distinction-making process described by the partition join operation. In terms
of the lattice of set partitions, such a "Type 1" process moves up in the
lattice to more refined partitions.\footnote{The usual notion of refinement of
partitions, i.e., $\pi=\left\{  B\right\}  $ is \textit{more (or equally)
refined} than $\sigma=\left\{  C\right\}  $, denoted $\sigma\preceq\pi$, if
for each $B\in\pi$, there is a $C\in\sigma$ such that $B\subseteq C$,
\textit{is} just the inclusion relation on distinctions, i.e., $\sigma
\preceq\pi$ iff $\operatorname*{dit}\left(  \sigma\right)  \subseteq
\operatorname*{dit}\left(  \pi\right)  $, so moving up in the ordering means
making more distinctions.} This means in QM/sets that a "Type 2" evolution
would be a distinction-preserving process that, as it were, moves horizontally
in the lattice of partitions.%

\begin{center}
\includegraphics[
height=1.599in,
width=3.819in
]%
{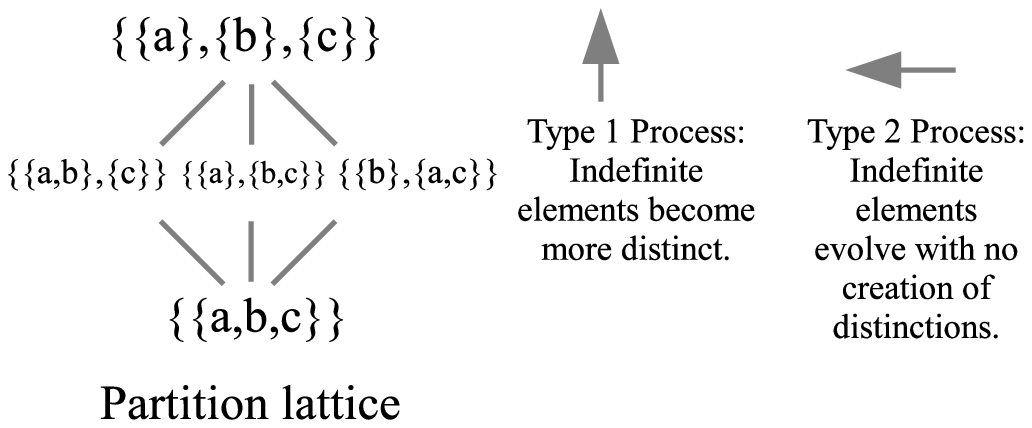}%
\end{center}

\begin{center}
Figure 8: "Type 1" distinction-making and "Type 2" distinction-preserving
processes in QM/sets
\end{center}

A linear transformation $\wp\left(  U\right)  \rightarrow\wp\left(  U\right)
$ that keeps distinct vectors distinct (i.e., preserves distinctions) is just
a non-singular transformation.\footnote{Thus the gates in quantum computing
over $%
\mathbb{Z}
_{2}$ are the non-singular linear transformations (\cite{schum:modal} and
\cite{ell:qmoversets}).} This means that a Type 2 process in full QM should be
a process that preserves the degree of distinctness and indistinctness. Given
two normalized quantum states $\psi$ and $\varphi$, the brackets $\left\langle
\psi|\varphi\right\rangle $ can be interpreted as the degree of indistinctness
(or identification) of the states with the extreme values of $\left\langle
\psi|\varphi\right\rangle =1$ for full indistinctness or identification, i.e.,
$\psi=\varphi$, and $\left\langle \psi|\varphi\right\rangle =0$ for zero
indistinctness, i.e., the full distinctness of orthogonality. Hence under this
partitional approach to understanding or making sense of QM, the
distinction-preserving Type 2 processes are the ones that preserve the degree
of indistinctness $\left\langle \psi|\varphi\right\rangle $, i.e., the unitary
transformations. Thus by this approach, we can philosophically derive the
mathematical form of the indistinctness-preserving transformations in full QM.
Moreover, by Stone's Theorem in mathematics, those unitary transformations are
given by (the complex exponential of) a Hermitian operator $H$ that would
describe how the "substance" changes to preserve indistinctness.\footnote{See
the previous Figure 3 on how "substance" is constant or conserved in the
partition lattice regardless of whether the changes preserve distinctions
("Type 2" processes--imaged in Figure 8 as lateral moves in the lattice) or
create distinctions ("Type 1" processes--imaged as upward moves in the
lattice). Also the "Type 1" processes that create distinctions will destroy or
wipe out distinctions in another incompatible basis.}

The partitional follow-the-math approach to understanding QM shows that the
mathematical form of full QM is a type of logical probability system expressed
in terms of vector spaces rather than sets (the usual setting for logic and
the Laplace-Boole logical probability theory). But at some point this
mathematical metaphysics has to pass the torch to physics to give empirical
content to the formalism. As previously noted, Heisenberg tells us that
"substance" is energy, but this "follow the partitional math" approach to
making sense out of QM of course cannot derive that physical fact nor the form
of the operator $H$ ("H" as in Hamiltonian) for a physical system.

There has been much concern and mystery in the foundations of quantum
mechanics about the Type 1 measurement processes--in contrast to the clarity
of describing the Type 2 processes as unitary transformations. That is another
old question. We have seen that this question can be answered at the
\textit{mathematical} level by the partition join operation. The mathematical
description of a measurement in QM/sets as the join of compatible
set-partitions lifts to the mathematical description of a measurement in full
QM as the join of compatible vector-space partitions--where one partition
describes the distinction-making measurement apparatus representing the
observable being measured and the other partition is the indiscrete (i.e., one
block) partition representing the given pure state to be measured or, better,
to be decohered.\footnote{See \cite{ell:qmoversets} for the development of
this using density matrices. This partition mathematical description of the
Type 1 measurement process does not give the physical characterization of a
distinction-creating measurement apparatus anymore than this approach gives
the physical characterization of the distinction-preserving Type 2 unitary
processes.} Using the respective partition join operations, a complete set of
compatible attributes (CSCA) gives a non-degenerate measurement in QM/sets,
and a complete set of commuting operators (CSCO) gives a non-degenerate
measurement in full QM. Thus the clear and distinct distinction between "Type
1" distinction-making and "Type 2" distinction-preserving processes in the
pedagogical model of QM/sets makes sense of the von Neumann Type 1
distinction-making measurements and Type 2 distinction-preserving unitary
transformations in full QM.

\section{Final remarks}

There are two meta-physical visions of reality suggested by classical physics
(objectively definite reality) and by quantum physics (objectively indefinite
reality). The problem of interpreting QM is essentially the problem of making
sense out of the notion of objective indefiniteness. Our sense-making strategy
was to follow the lead of the mathematics.

The definiteness of classical physics is associated with the notion of a
subset and is logically expressed in the classical Boolean logic of subsets.
The indefiniteness of quantum physics is associated with the notion of a
quotient set, equivalence relation, or partition, and the corresponding logic
is the recently developed logic of partitions \cite{ell:partitions}. Moreover,
those associated notions of subsets and quotient sets are
category-theoretically dual to one another, so \textit{from that viewpoint},
those are \textit{the only} two possible frameworks to describe reality.
Common sense and classical physics assumes the objectively definite type of
reality, but quantum physics strongly indicates an objectively indefinite
reality at the quantum level. Hence our approach to interpreting quantum
mechanics is not flights of fantasy (e.g., about many worlds or realms of
hidden variables) but is trying to make sense out of objective indefiniteness.

Our sense-making strategy was implemented by developing the mathematics and
logic of partitions at the connected conceptual levels of sets and vector
spaces. Set concepts are transported to complex vector spaces to yield the
mathematical machinery of full QM, and the complex vector space concepts of
full QM are transported to the set-like vector spaces over $%
\mathbb{Z}
_{2}$ to yield the rather fulsome pedagogical model of quantum mechanics over
sets or QM/sets.

In this manner, we have tried to use partition concepts to make sense of
objective indefiniteness and thus to interpret quantum mechanics.

\end{document}